\documentclass[pra,reprint]{revtex4-2}
\usepackage{graphicx,url,hyperref,bm,makecell,tabularray,amsmath,braket,enumitem}
\usepackage[dvipsnames]{xcolor}
\usepackage{tikz,stix}
\renewcommand{\vec}[1]{\bm{#1}}

\begin{document}

\title{Measured dynamics of an XXZ quantum simulator
  in a highly symmetrical double--ringed geometry}

\author{D. J.\ Papoular}
\email[Electronic address: ]{david.papoular@cyu.fr}
\affiliation{Laboratoire de Physique Th\'eorique et Mod\'elisation,
  UMR 8089 CNRS \& CY Cergy Paris Université,
  95302 Cergy--Pontoise,
  France}
\date{\today}

\begin{abstract}
  We theoretically identify observable
  consequences of spatial and spin symmetries on the
  dynamics of a small XXZ quantum simulator.
  Our proposed protocol
  relies on the choice of suitable initial states, and
  involves the measurement scheme whose experimental implementation is the simplest.
  We analyze a system of $N=2n=6$ to $12$ particles,
  trapped
  in a planar geometry comprised of two rings
  which exhibits point group symmetry $D_{nh}$.
  The particles represent effective spins
  whose interaction is described by
  the XXZ or Heisenberg Hamiltonian.
  The system is prepared in an initial state
  which is  sitewise--factorized and invariant under all spatial symmetries,
  it evolves for a given time, after which
  the $z$--components of all $N$ spins are measured.
  We show that symmetries dictate \textit{(i)} the
  qualitative behaviour of the measurement probabilities as a function of
  the evolution time, and
  \textit{(ii)} the number
  of measurement results with different probabilities.
  We highlight the role of a twofold rotation of  all spins.
  We also demonstrate that, in larger systems, the collapse of the initial state
  may be observed.
\end{abstract}

\maketitle


\begin{figure*}
  \includegraphics[width=\textwidth]
  {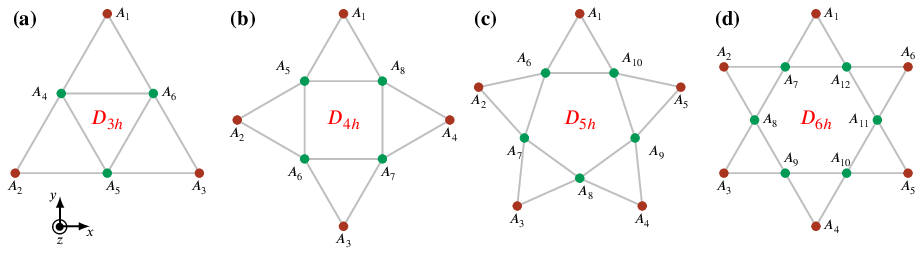}
  \caption{\label{fig:allgeoms}
    The four considered geometries,
    each comprised of an even number $N=2n$ of particles
    with
    $n=3$ {(a)}, $4$ {(b)}, $5$ {(c)}, and $6$ {(d)},
    trapped in the $(x,y)$ plane at
    the sites $(A_i)_{1\leq i\leq N}$, 
    with  one particle per site.
    Each geometry exhibits point group symmetry $D_{nh}$,
    but no translational invariance.
    It involves a double--ring structure, with $n$
    particles on the outer ring (sites $1\leq i\leq n$, shown in red)
    and $n$ particles on the inner ring
    (sites $n+1\leq i\leq N$, shown in green).
    Neighboring atoms linked by gray segments are equidistant.
  }
\end{figure*}

\section{Introduction}

The symmetries of a system
provide a wealth of information concerning it without the need for an explicit solution.
In the context of quantum mechanics, they have  been exploited
to explain the degeneracies of the energy spectrum,
to analyze energy level splittings due to
perturbations, and to establish selection rules
\cite[ch.~1]{heine:Pergamon1960}.
They have been instrumental in the interpretation
of the energy spectra of atoms, molecules, and crystals
\cite[chs.~6--8]{tinkham:McGrawHill1964}.
Their impact on quantum dynamics has long been
investigated (see e.g.\ Ref.~\cite[Sec.~11.5]{FLP3:BB2010}).

There is a  fundamental interest in studying
spin systems whose sites exhibit high spatial symmetry.
These systems 
exhibit two types
of symmetries: firstly, the discrete symmetries affecting the positions
of the particles, and secondly,
the continuous or discrete symmetries affecting their spins.
Symmetries of one type may be applied to the system independently
of those of the other. The set of all symmetries, comprised of
symmetries of both types and of their products,
is a spin point group \cite{brinkman:ProcRSocA1966,litvin:Physica1974}.
Such symmetry groups are currently being investigated in the context
of condensed matter physics \cite{xiao:PRX2024,schiff:SciPostPhys2025}

These highly--symmetrical spin systems may be realized, manipulated,
and measured,
owing to recent experimental
progress in the field of quantum simulation 
\cite{gross:Science2017,browaeys:NatPhys2020}.
There, conceptually simple systems are constructed
using $N$ trapped particles which may be e.g.\ magnetic
atoms, alkali atoms in Rydberg states, or polar molecules
\cite{paz:PRL2013,kaufman:NatPhys2021},
including the minimal ingredients yielding the sought effect.
The particles may be confined in
well--controlled individual traps arranged
in arbitrary geometries 
\cite{barredo:Science2016,barredo:Nature2018,
  kaufman:NatPhys2021,scholl:Nature2021,ravon:PRL2023,
  bornet:arXiv2026}.
Each particle represents a two--level system.
The system may be prepared in an initial $N$--particle state
which is an arbitrary tensor product of
individual spin states by applying local electromagnetic pulses \cite[Sec.~1.5.2]{nielsen:CUP2000}.
The pair--wise interaction between the particles may be tailored to simulate a spin system
described by the Heisenberg or XXZ Hamiltonians
\cite{yan:Nature2013,geier:Science2021,scholl:PRXQ2022,nguyen:PRX2018}.
The system evolves under this Hamiltonian for a given time.
Then, the final state of the $N$ particles may be measured
using optical methods
\cite{barredo:Science2016,barredo:Nature2018,christakis:Nature2023,machu:arXiv2025}.
Multiple realizations of the experiment 
provide the measurement probabilities.
Such schemes have already allowed for the experimental characterization  
of e.g.\ 2D antiferromagnets \cite{scholl:Nature2021,ebadi:Nature2021}.

Observable signatures of the high symmetry of the Hamiltonian of a quantum system
are subtle to identify. This is due to the fact that observed signals
result from two consecutive steps:
the quantum evolution, followed by a quantum measurement.
The evolution is piloted by the Hamiltonian and, hence,
enforces the conservation laws corresponding to all its symmetries.
The measurement of a given  observable is characterized
by a specific basis of
the $N$--particle Hilbert space,
comprised of its eigenstates, which are the possible
measurement results \cite[Sec.~VIII--2]{messiah1:NorthHolland1962}.
For the measurement to enforce the conservation laws of the Hamiltonian,
this basis
should  be comprised of  symmetry--adapted linear combinations
\cite[ch.~6]{cotton:Wiley1990}, i.e.\
states each transforming under an 
irreducible representation of the symmetry group
\cite[\S 94]{landau3:BH1977}. 
However,  the measurement of arbitrary observables
is currently out of experimental reach
\cite[Sec.~4.5.4]{nielsen:CUP2000}.
The accessible observables are
determined by experimental requirements,
and their eigenstates are not adapted to the larger symmetry groups.
Thus, the probability
amplitudes characterizing the measurement result from the interference between
components of the wavefunction with different symmetry properties.
This interference is similar to that which affects
the measurement of a dressed two--level system,
yielding Rabi oscillations in the probability amplitudes
as a function of time \cite[Sec.~IV.C.3]{CDL1:Wiley2020}.
If the interference is not avoided, the measurement probabilities exhibit
no clear signature of the symmetries of the Hamiltonian.
One way of avoiding it,
already considered in the literature (see e.g.\
Ref.~\cite{musolino:PRL2024,antonic:PRB2025}),
is to measure well--chosen observables
compatible with a specific subset of operations in the symmetry group.

In this article, we propose an alternative approach
applicable to spin systems with high spatial symmetry.
We show that the interference may be avoided by  
selecting suitable initial  states, all easily prepared,
which are invariant under all spatial symmetries.
Their quantum evolution may  be probed by the measurement scheme
whose experimental implementation  is the most straightforward, namely,
the simultaneous measurement of the $z$--components of all  spins.
We identify signatures of the spatial and spin symmetries of the Hamiltonian
in the time dependence of the resulting measurement probabilities.

We present our approach on the case of a system of effective
spins represented by particles which may be either bosonic or fermionic,
confined in a planar geometry. 
In order for our analysis to be relevant to current experiments,
we keep their number $N$ relatively small,
and consider up to a dozen particles. Two--dimensional systems comprised of
so few particles  would realize only a poor approximation
of lattice translational invariance. Therefore, we
do not seek to simulate a crystal, but rather a highly symmetrical molecule.
Thus, the spatial symmetries of the considered systems make up a point group
\cite[\S 93]{landau3:BH1977}, specifically $D_{nh}$ with $n=3$, $4$, $5$, or $6$
and $N=2n$. The considered geometries, represented in Fig.~\ref{fig:allgeoms},
are each comprised of two concentric rings involving $n$ particles.
Compared to geometries involving a single ring,
the double--ring structure allows for a greater variety of
experimentally accessible initial states,
a flexibility we shall explicitly make use of.

The dynamics of the $N$--particle system we are considering is governed
either by the Heisenberg Hamiltonian, or by the more general XXZ Hamiltonian.
We show that our choice of initial states constrains the quantum dynamics
of the $N$--particle system to occur within the subspace of the full Hilbert space
comprised of the states which are invariant under all spatial symmetries.
The dimension of this subspace plays a key role, and its value has two
observable consequences on the measurement probabilities.
Firstly, this dimension determines the qualitative behavior
of the probabilities as a function of the evolution time: these may
be constant, or oscillate sinusoidally, or undergo an aperiodic evolution.
Secondly, symmetries cause many of the possible measurement results to have
the same probability at all times, and the number of possible measurement
results with different probabilities
is equal to the dimension of the subspace within which the evolution occurs.

In the specific case of the XXZ Hamiltonian,
we analyze the role of the twofold rotations of the $N$ spins about an axis 
in the horizontal plane (plane of Fig.~\ref{fig:allgeoms}), and explain
how the conservation of the corresponding parity may be stringently tested
on smaller systems, comprised of $N=6$ spins.
Finally, we identify an effect occurring in larger systems comprised
of e.g.\ $N=12$ spins, namely, the collapse of the component of the initial
state with zero total spin projection.

\subsection*{Outline}
This article is organized as follows.
In Sec.~\ref{sec:system-protocol}, we introduce 
the considered $N$--particle system and its Hamiltonian, and
put forward  our proposed protocol. 
In Sec.~\ref{sec:symmetries_initstates}, we describe the spatial and
spin symmetries of the system, and select initial states allowing
for their investigation.
In Sec.~\ref{sec:two_observable_consequences},
we identify two observable consequences of symmetry on the time dependence
of the measurement probabilities: their qualitative behavior, and
the number of measurement results with different probabilities.
In Sec.~\ref{sec:XXZ_xi}, we analyze specific cases within experimental reach.
We highlight the role of the twofold rotation of the $N$ spins, which may
be probed on smaller systems ($N=6$) described by the XXZ Hamiltonian.
We also demonstrate that larger systems ($N=12$) allow for the observation
of the collapse of the initial state.
Finally, we conclude in Sec.~\ref{sec:Conclusion}.

\section{ \label{sec:system-protocol}
  Considered system and protocol}

\subsection{System and effective spin Hamiltonian}

We consider  the four geometries of Fig.~\ref{fig:allgeoms}(a-d),
respectively comprised of $N=6$, $8$, $10$, and $12$ particles trapped at the sites
$(A_i)_{1\leq i\leq N}$ whose positions are $(\vec{r}_i)_{1\leq i\leq N}$,
with one particle per site.
These geometries are planar, with all sites lying in the $(xy)$ plane.
They consist of two rings, each comprised of $n$ sites.
The sites on the outer ($1\leq i\leq n$) and inner ($n+1\leq i\leq N$)
rings appear in red and green, respectively, in Fig.~\ref{fig:allgeoms}.
The neighboring sites linked by gray segments  are equidistant.

We neglect the spatial motion of the particles within the traps.
Each particle $i$ behaves as a two--level system, whose two accessible quantum states
we call $\ket{\uparrow^z_i}$ and $\ket{\downarrow^z_i}$.
The particles exhibit pairwise interaction, whose strength
depends on the distance $r_{ij}=|\vec{r}_j-\vec{r}_i|$ between the
sites $A_i$ and $A_j$
through the power law $1/r_{ij}^\alpha$ with $\alpha>0$, with $\vec{r}_i$
being the position of the site $A_i$. The $N$--particle Hamiltonian $H$
involves interactions between all pairs of particles but no intersite
tunneling:
\begin{equation}
  \label{eq:XXZham}
  H=
  \frac{1}{2}\sum_{i\neq j}
  \left(\frac{a}{r_{ij}}\right)^\alpha
  \left[
    J(\sigma_i^x\sigma_j^x+\sigma_i^y\sigma_j^y)
    +
    J_z\, \sigma_i^z \sigma_j^z
  \right]
  \ ,
\end{equation}
where $a$ is the nearest--neighbor distance
($a=r_{14}$, $r_{15}$, $r_{16}$, and $r_{17}$, respectively,
for the four geometries of Fig.~\ref{fig:allgeoms}).
The Pauli operators  $\vec{\sigma}_i=(\sigma^x_i,\sigma^y_i,\sigma^z_i$)
represent the two--level system trapped at $A_i$.
The coefficients $J$ and $J_z$ are constant energies.
The Hamiltonian $H$ is the Heisenberg Hamiltonian $H_\mathrm{H}$ or
the XXZ Hamiltonian $H_\mathrm{XXZ}$
depending on whether
$J=J_z$ or $J\neq J_z$.
We have performed all  the numerical calculations reported in this paper
using $\alpha=6$, $J>0$, and $-3<J_z/J<3$, matching the proposal of
Ref.~\cite{nguyen:PRX2018} involving circular Rydberg atoms.
However, these choices are not critical, and our analysis may be extended to
other experimental realizations
of the 2D Heisenberg and XXZ models
(see e.g.\ Refs.~\cite{paz:PRL2013,scholl:PRXQ2022,christakis:Nature2023}).

Regardless of the bosonic or fermionic nature of the actual trapped
particles, the simultaneous assumptions
of a single particle per site and no intersite tunneling allow us to describe
the system in terms of $N$ distinguishable effective  spins--$1/2$, represented
by the operators $\vec{s}_i=\hbar\vec{\sigma}_i/2$.
\subsection{ \label{sec:protocol}
  Protocol yielding
  time--dependent probabilities}
We analyze the following protocol ($\mathcal{P}$).
The system is initially prepared in an
$N$--particle state $\ket{\psi_0}$.
It evolves under the Hamiltonian $H$ of Eq.~\eqref{eq:XXZham}
for the  duration $t$,
giving rise to the quantum state
$\ket{\psi(t)}=e^{-iHt/\hbar}\ket{\psi_0}$.
At time $t$, we measure the observable $s_i^z$ for each of the $N$
spins. This yields one of
$2^N$ possible results, namely, the $N$--particle state
$\ket{c_f}=\ket{\mu_1,\ldots,\mu_N}$
where the spin at site
$A_i$ is in the state $\ket{\mu_i}=\ket{\uparrow^z_i}$
or $\ket{\downarrow^z_i}$. The $2^N$ configurations
$\ket{c_f}$
make up the basis  $\mathcal{C}=(\ket{c_f})_{1\leq f\leq 2^{N}}$
of the full Hilbert space $\mathcal{H}$.
(The ordering of the basis states $\ket{c_f}$
is detailed in Appendix \ref{sec:config_numbering}).

Such a measurement 
is experimentally accessible
(see Ref.~\cite{machu:arXiv2025} 
for a recent demonstration with circular Rydberg atoms).
Multiple repetitions of the sequence, with the same
initial state $\ket{\psi_0}$ and duration $t$, give
access to
the probabilities $p_f(t)=|\braket{c_f|\psi(t)}|^2$
for the measurement to yield the result $\ket{c_f}$.
These probabilities depend on the chosen time $t$.

The goal of the present work is to identify
experimentally accessible properties of the time--dependent $p_f(t)$'s,
holding for specific initial states $\ket{\psi_0}$,
which exhibit signatures of the spatial and spin symmetries of the
$N$--particle system.

\section{ \label{sec:symmetries_initstates}
  Symmetries of the Hamiltonian and choice of the initial states}

\subsection{\label{sec:sym_space_spin} Spin--point group comprising spatial and spin symmetries}

We present the group of all symmetries of the Hamiltonian $H$
represented by unitary operators
as a spin--point group \cite{brinkman:ProcRSocA1966,litvin:Physica1974}
$G=G^\mathrm{spatial}\times G^\mathrm{spin}$, 
which is the direct product of the group $G^\mathrm{spatial}$
acting on the positions
while leaving the internal states of the effective spins unchanged,
and the group $G^\mathrm{spin}$ acting on the internal states
while leaving the positions unchanged.

\emph{Spatial symmetries ---}
The trapping geometries of Fig.~\ref{fig:allgeoms}, involving at most
12 sites, do not exhibit translational invariance. Accordingly, their spatial
symmetry properties are those of a molecule (rather than of a crystal).
The spatial symmetry group of the geometry involving $N=2n$ particles is
the point group $G^\mathrm{spatial}=D_{nh}$ \cite[\S 93]{landau3:BH1977}.
It  contains $4n$ elements, all obtained as products of
the rotation of order $n$ about the axis $\vec{z}$,
the rotation  of order $2$ about the axis $\vec{y}$,
and the reflection  in the horizontal plane $(Oxy)$.
Each element of $D_{nh}$ 
is characterized by
a permutation $\phi$ mapping the $N$ sites
$(A_{i})_{1\leq i\leq N}$ onto $(A_{\phi(i)})_{1\leq i\leq N}$.
Then, 
the unitary operator $U_\phi$
representing this element, which
acts  on the Hilbert space $\mathcal{H}$,
maps each configuration
$\ket{c_f}=\ket{\mu_1,\ldots,\mu_N}$ in the basis $\mathcal{C}$
onto the configuration
$U_\phi\ket{c_f}=\ket{\mu_{\phi^{-1}(1)},\ldots,\mu_{\phi^{-1}(N)}}$, also in $\mathcal{C}$.

\emph{Spin symmetries ---}
The spin symmetry group $G^\mathrm{spin}$
depends on the values of $J$ and $J_z$
in Eq.~(\ref{eq:XXZham}).
If $J=J_z$ (i.e.\ $H=H_\mathrm{H}$),
$G^\mathrm{spin}=K_h$ is the group of
complete spherical symmetry \cite[\S 98]{landau3:BH1977},
including
spin rotations through any angle about any axis in
three--dimensional space.
If $J\neq J_z$ (i.e.\ $H=H_\mathrm{XXZ}$),
$G^\mathrm{spin}$ is the smaller group
$D_{\infty h}$, including
spin rotations through any angle about the $z$ axis, and
spin rotations through angle $\pi$ about any horizontal axis.
In both cases, each element $g$ in $G^\mathrm{spin}$
is represented by the unitary operator
$U_g=u_g^{(1)}\cdots u_g^{(N)}$, where $u_g^{(i)}$
acts on the state of the spin at site $A_i$ 
in the same way as it would act on a true spin--$1/2$
\cite[Secs.~XIII.19 \& XV.10]{messiah2:NorthHolland1962}.

\subsection{ \label{sec:conservationlaws_measurement}
  Interplay between conservation laws and measurement}

The symmetries of Sec.~\ref{sec:sym_space_spin} yield conservation laws
which hold at all times during the evolution described by the Schr\"odinger
equation, i.e.\ up to just before the measurement is performed.
We point out two of them which are valid for both the Heisenberg and the XXZ Hamiltonians.
Firstly, the presence in $G^\mathrm{spin}$ of all rotations about the axis $\vec{z}$
yields the conservation of the total spin projection operator $S_z=\sum_{i=1}^N s_i^z$
\cite[\S 26]{landau3:BH1977}.
Secondly, if the initial $N$--particle state $\ket{\psi_0}$ transforms according
to a given irreducible representation $\rho$ of the spatial symmetry group $D_{nh}$,
then so does the $N$--particle state $\ket{\psi(t)}$ \cite[\S 97]{landau3:BH1977}.
Any initial state $\ket{\psi_0}$ is a linear superposition of
components $\ket{\psi_0^{\rho,M}}$, each of which transforms according to the irreducible
representation $\rho$ of $D_{nh}$ and is an eigenstate of $S_z$ with eigenvalue $\hbar M$,
where the total spin projection $M$ is an integer such that
$-n\leq M\leq n$.
Owing to the two conservation laws stated above,
these components evolve independently from one another
up to just before the measurement, 
and $\ket{\psi(t)}=\sum_{\rho,M}\ket{\psi^{\rho,M}(t)}$ with
$\ket{\psi^{\rho,M}(t)}=e^{-iHt/\hbar}\ket{\psi_0^{\rho,M}}$.

We now discuss  the impact of these two conservation laws on
the probability amplitude $\braket{c_f|\psi(t)}$ for
the measurement performed at time $t$ to yield the result
$\ket{c_f}=\ket{\mu_1,\ldots,\mu_N}$, which is
an $N$--particle state in the basis $\mathcal{C}$.
Firstly, we consider the total spin projection.
The state $\ket{c_f}=\ket{c_f^{M_f}}$
is an eigenstate of the operator $S_z$ with the eigenvalue $\hbar M_f$, where the total spin
projection
$M_f=\sum_{i=1}^N\mu_i$ and $\mu_i=\pm 1/2$ according to whether
$\ket{\mu_i}=\ket{\uparrow_i^z}$ or $\ket{\downarrow_i^z}$.
Therefore, only the components $\ket{\psi^{\rho,M_f}(t)}$  with $M=M_f$
contribute to $\braket{c_f|\psi(t)}$,

Secondly, we turn to the irreducible representations $\rho$ of the
spatial symmetry group $D_{nh}$.
Most states $\ket{c_f}$ in the basis $\mathcal{C}$ do not transform
under a specific irreducible representation $\rho$. Instead,
they are superpositions
$\ket{c_f}=\sum_\rho \ket{c_f^\rho}$ of multiple
components $\ket{c_f^\rho}$, each transforming under
a given  representation $\rho$. 
Then, the probability amplitude $\braket{c_f|\psi(t)}$
reads:
\begin{equation}
  \label{eq:cfpsi_sumrho}
  \braket{c_f|\psi(t)}=\sum_\rho\braket{c_f^\rho|\psi^{\rho,M_f}(t)}
  \ .
\end{equation}  
Unless 
the sum in Eq.~\eqref{eq:cfpsi_sumrho} reduces to a single term,
the measurement causes interference between the wavefunction components $\ket{\psi^{\rho,M_f}(t)}$
with the same total spin projection $M_f$, but transforming under different
irreducible representations $\rho$ of $D_{nh}$, so that
the probabilities $p_f(t)=|\braket{c_f|\psi(t)}|^2$ exhibit no clear
signature of the spatial symmetry group $D_{nh}$.

\subsection{ \label{sec:initstates}
  The considered initial states}
We avoid the interference identified in 
Sec.~\ref{sec:conservationlaws_measurement} by selecting initial $N$--particle
states  $\ket{\psi_0}$ which transform under a given irreducible
representation $\rho_0$ of the spatial symmetry group $D_{nh}$. Then, each component
$\ket{\psi^M(t)}=\ket{\psi^{\rho_0,M}(t)}$ of $\ket{\psi(t)}$ with total spin projection $M$
also transforms under $\rho_0$, and the sum of Eq.~(\ref{eq:cfpsi_sumrho})
reduces to a single term, $\braket{c_f|\psi(t)}=\braket{c_f^{\rho_0}|\psi^{\rho_0,M_f}(t)}$.
In this equality, the representation $\rho_0$ is the one under which the initial state
$\ket{\psi_0}$ transforms, whereas the total spin projection $M_f$ is that of the
measurement result $\ket{c_f}$.

This situation may be achieved experimentally by selecting initial
states of the form $\ket{\psi_0}=\ket{\chi_{\vec{u},\vec{v}}}$ defined
as follows:
\begin{equation}
  \label{eq:def_chi_uv}
  \ket{\chi_{\vec{u},\vec{v}}}=
  \ket{
    \uparrow_1^{\vec{u}},\ldots,\uparrow_n^{\vec{u}},
    \uparrow_{n+1}^{\vec{v}},\ldots,\uparrow_N^{\vec{v}}
  }
  \ ,
\end{equation}
where all $n$ spins on the sites $A_i$ of the outer ring
($i=1$ to $n$)
are in the same single--particle state $\ket{\uparrow_i^{\vec{u}}}$,
and all $n$ spins on the sites of the inner ring
($i=n+1$ to $N$) 
are in the same state $\ket{\uparrow_i^{\vec{v}}}$.
The real unit vectors $\vec{u}$ and $\vec{v}$ represent two directions
on the Bloch sphere \cite[Sec.~1.2]{nielsen:CUP2000}, and
for $\vec{w}=\vec{u}$ or $\vec{v}$, the state
$\ket{\uparrow_i^{\vec{w}}}=
\cos(\theta/2)e^{-i\phi/2}\ket{\uparrow_i^z}
+
\sin(\theta/2)e^{i\phi/2}\ket{\downarrow_i^z}$,
with $(\theta,\phi)$ being the spherical coordinates of $\vec{w}$.

The $N$--particle state $\ket{\chi_{\vec{u},\vec{v}}}$ is a tensor product
of $N$ single--particle states, hence, it may be prepared experimentally,
e.g.\ starting from the polarized state
$\ket{\uparrow^z_1,\ldots,\uparrow^z_{N}}$ and applying
electromagnetic pulses to the individual  spins
\cite[Sec.~1.5.2]{nielsen:CUP2000}.

The spatial symmetries in the group $D_{nh}$, acting on Hilbert space as the operators
$U_\phi$ of Sec.~\ref{sec:sym_space_spin},
permute the states of the
$n$ spins on the sites of the outer ring among themselves, and those of the $n$
spins on the inner ring among themselves.  
Therefore, the state $\ket{\chi_{\vec{u},\vec{v}}}$ is invariant under all
spatial symmetries. This amounts to stating that it transforms under
the unit representation $\rho_1$ of $D_{nh}$, which is
irreducible \cite[\S 94]{landau3:BH1977}.

To sum up, the $N$--particle states
$\ket{\chi_{\vec{u},\vec{v}}}$ of Eq.~(\ref{eq:def_chi_uv}), which are experimentally accessible,
transform under the unit representation $\rho_1$ of the spatial symmetry group $D_{nh}$.
Their
being indexed by two independent directions $\vec{u}$ and $\vec{v}$
follows from the presence of two rings in the geometries of
Fig.~\ref{fig:allgeoms}. This family of states is sufficiently large 
to allow for the observation of various qualitative behaviors,
discussed below,
for the time dependence of the measurement probabilities.

\emph{States with maximal total spin modulus ---}
As a special case of Eq.~\eqref{eq:def_chi_uv}, we first consider the state
$\ket{\xi_{\vec{u}}}=\ket{\chi_{\vec{u},\vec{u}}}=\ket{\uparrow_1^{\vec{u}},\ldots,\uparrow_N^{\vec{u}}}$,
describing $N=2n$ particles all in the same single--particle state $\ket{\uparrow_i^{\vec{u}}}$,
for a given direction $\vec{u}$ on the Bloch sphere with spherical coordinates $(\theta,\phi)$.
The state $\ket{\xi_{\vec{u}}}$ is an eigenstate of the squared total spin
operator $\vec{S}^2=(\vec{s}_1+\ldots+\vec{s}_N)^2$ with the eigenvalue
$\hbar^2 S(S+1)$, the total spin modulus $S=n$ being maximal.
Hence, $\ket{\xi_{\vec{u}}}$  is also an eigenstate of the Heisenberg Hamiltonian $H_\mathrm{H}$
\cite[ch.~33]{ashcroft:Saunders1976}, and its evolution under $H_\mathrm{H}$
leads to measurement probabilities that are all
constant:
$p_f^M=|\braket{c_f^M|\psi(t)}|^2=\cos^{N+2M}(\theta/2)\sin^{N-2M}(\theta/2)$
for 
all states $\ket{c_f^M}$ in the basis
$\mathcal{C}$ with total spin projection $M$.

The protocol $\mathcal{P}$ of Sec.~\ref{sec:protocol}
yields time--dependent probabilities $p_f^M(t)$ for 
three possible combinations of initial states and Hamiltonians:
\textit{(i)} the
initial state $\ket{\psi_0}=\ket{\xi_{\vec{u}}}$ evolving
under the XXZ Hamiltonian with $J_z\neq J$;
or the initial state $\ket{\psi_0}=\ket{\chi_{\vec{u},\vec{v}}}$ with
$\vec{u}\neq\vec{v}$ evolving under
\textit{(ii)} the XXZ Hamiltonian or \textit{(iii)} the
Heisenberg Hamiltonian.
These three cases are respectively considered in
Secs.~\ref{sec:XXZ_xi}, \ref{sec:XXZ_twofoldspinrotation},
and \ref{sec:Heisenberg_chi_collapse} below.
Their discussion first requires the introduction of
two observable consequences of the spatial and spin symmetries
onto the time dependence of the measurement probabilities $p_f^M(t)$.

\section{ \label{sec:two_observable_consequences}
  Observable consequences of symmetry
  on the time--dependent probabilities
}

\begin{table}
  \begin{tblr}{|c|c|c|c|c|}
    \hline
    & $N=6$ & $N=8$ & $N=10$ & $N=12$
    \\
    \hline
    \hline
    $(\rho_1,M=\pm 6)$ &
    \SetCell{bg=lightgray}N/A & \SetCell{bg=lightgray}N/A & \SetCell{bg=lightgray}N/A  &  $1$
    \\
    \hline
    $(\rho_1,M=\pm 5)$ &
    \SetCell{bg=lightgray}N/A &  \SetCell{bg=lightgray}N/A & $1$  & $2$
    \\
    \hline
    $(\rho_1,M=\pm 4)$ &
    \SetCell{bg=lightgray}N/A &  $1$ & $2$  & $9$
    \\
    \hline
    $(\rho_1,M=\pm 3)$ &
    $1$ &  $2$ & $7$  & $24$
    \\
    \hline
    $(\rho_1,M=\pm 2)$ &
    $2$ &  $6$ & $16$ & $50$
    \\
    \hline
    $(\rho_1,M=\pm 1)$ &
    $4$ & $10$ & $26$ & $76$
    \\
    \hline
    $(\rho_1,M=0,\text{even}/Y^\mathrm{spin})$ &
    $3$ & $8$ & $16$ &  $48$
    \\
    \hline[dashed]
    $(\rho_1,M=0,\text{odd}/Y^\mathrm{spin})$ &
    $3$ & $5$ & $16$ & $42$
    \\
    \hline
\end{tblr}
\caption{ \label{tab:groupth_results}
  Dimensions  
  of the subspaces $(\rho_1,M)$,
  with $\rho_1$ being the unit representation of $D^\mathrm{spatial}_{nh}$,
  for the four geometries of Fig.~\ref{fig:allgeoms}.
  For $M=0$, the states transforming under $\rho_1$
  are further sorted in terms of their even or odd parity
  with respect to the operator $Y^\mathrm{spin}$, representing
  the rotation of all spins through angle $\pi$ about the axis $\vec{y}$.
  The non--applicable (N/A) cells with  $|M|>n$
  are shaded in gray.
  }
\end{table}

From this point on, we choose the initial state used in the protocol $(\mathcal{P})$
of Sec.~\ref{sec:protocol} to be of the form of Eq.~(\ref{eq:def_chi_uv}),
i.e.\ $\ket{\psi_0}=\ket{\chi_{\vec{u},\vec{v}}}$.
Under this assumption, we identify in Secs.~\ref{sec:qualitativebehavior} and
\ref{sec:equiv_results} below
two observable consequences of the spatial and spin symmetries
onto the time dependence of the measurement probabilities
$p_f^M(t)=|\braket{c_f^M|\psi(t)}|^2$, both of which
may be verified on current experimental setups.
\textit{(A)}
The first consequence concerns the qualitative behavior of the 
probabilities $p_f^M(t)$, which may be constant, oscillate sinusoidally,
or undergo an aperiodic evolution.
\textit{(B)} The second consequence is that many probabilities $p_f^M(t)$
are equal at all times.

Both of these consequences follow from 
the fact
that symmetries constrain the quantum evolution to occur within
a subspace whose dimension is  smaller than
the number of possible measurement results.
Indeed,
owing to the conservations laws
of Sec.~\ref{sec:conservationlaws_measurement},
the $N$--particle state
$\ket{\psi}=\sum_{M=-n}^{n}\ket{\psi^{\rho_1,M}}$
is a sum of components $\ket{\psi^{\rho_1,M}}$.
Each component evolves independently of the others,
within the subspace $(\rho_1,M)$
of Hilbert space comprised of all $N$--particle states which simultaneously
\textit{(i)} transform under
the representation $\rho_1$ of $D_{nh}$, and
\textit{(ii)} are eigenstates of
the operator $S_z$ with the total spin projection $M$.
Its dimension $\dim(\rho_1,M)$
is entirely determined by the symmetries
of the Hamiltonian, independently of the values of
the parameters $J$, $J_z$, and $\alpha$ entering Eq.~(\ref{eq:XXZham}).
The component $\ket{\psi^{\rho_1,M}(t)}$ determines the 
probabilities $p_f^M(t)=|\braket{c_f^{M}|\psi^{\rho_1,M}(t)}|^2$
for all $\binom{N}{n+M}$
measurement results $\ket{c_f^{M}}$ in the basis $\mathcal{C}$
with total spin projection $M$.

We calculate the dimension $\dim(\rho_1,M)$ of each subspace
$(\rho_1,M)$ by constructing the projector onto it
using well--established group--theoretical
methods \cite[\S 94]{landau3:BH1977}. Our results for all four geometries
of Fig.~\ref{fig:allgeoms} and all allowed values of $M$ are collected
in Table~\ref{tab:groupth_results}. They are noticeably smaller
than $\binom{N}{n+M}$
(except if the total spin projection satisfies $|M|=n$).

We now derive in turn both properties \textit{(A)} and \textit{(B)}
announced at
the beginning of the present section \ref{sec:two_observable_consequences}.

\subsection{ \label{sec:qualitativebehavior}
  Qualitative behavior of the measurement probabilities}

\begin{figure*}
  \includegraphics[width=\textwidth]
  {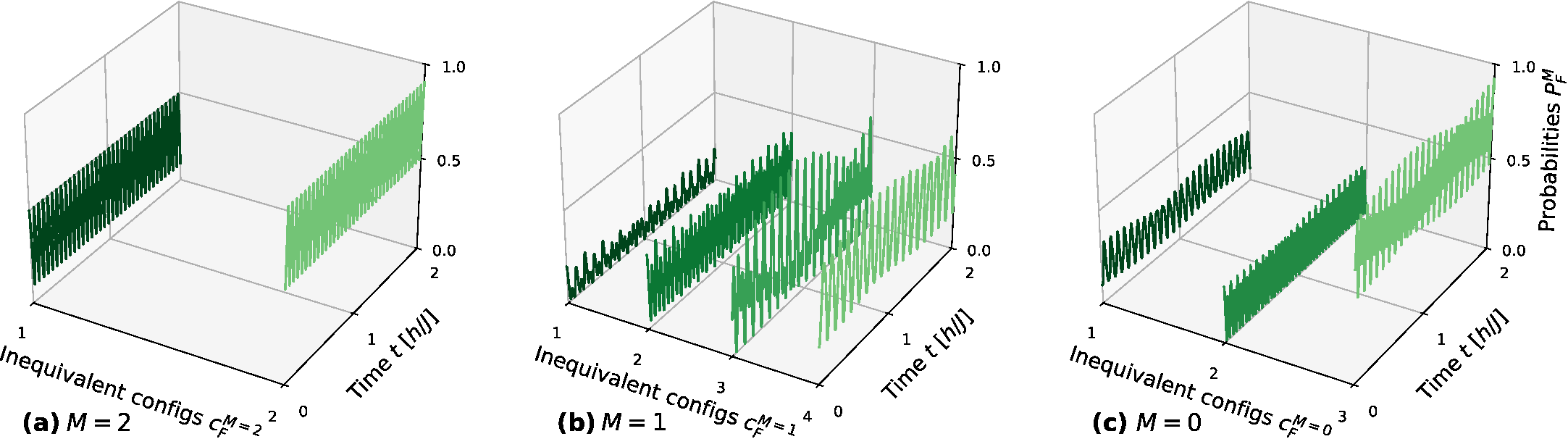}
  \includegraphics[width=\textwidth]
  {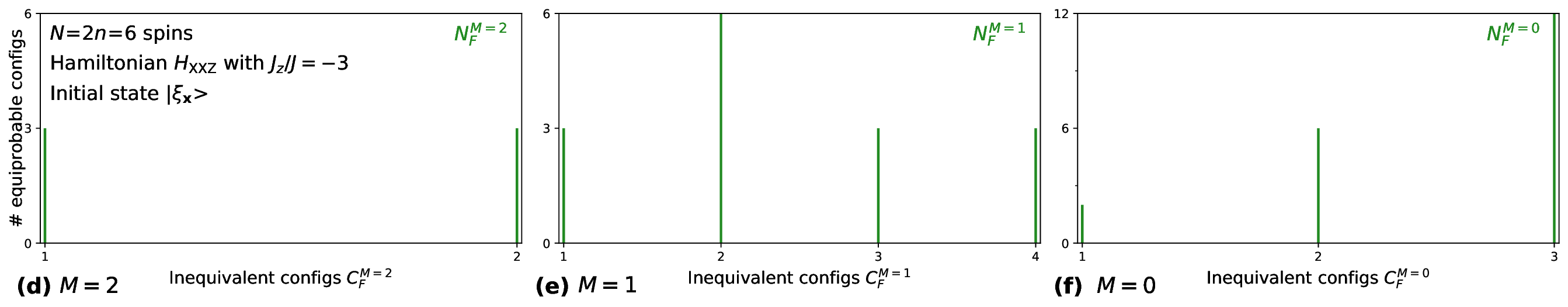}

  \caption{ \label{fig:probs_timedep_xit6}
    Panels (a--c):
    Time--dependent probabilities for the
    inequivalent measurement results $\ket{c^M_F}$
    with total spin projection $M$,
    for the 
    initial state $\ket{\psi_0}=\ket{\xi_\mathbf{x}}$ involving $N=6$ particles,
    evolving under the Hamiltonian
    $H_\mathrm{XXZ}$ with $J_z/J=-3$
    and $\alpha=6$,
    for $M=2$ (a), $1$ (b), and $0$ (c).
    For each value of $M$, we represent the quantities
    $P_F^M(t)=p_F^M(t) \: N_F^M/\|\psi_0^M\|^2$,
    as defined in Sec.~\ref{sec:XXZ_xi}.
    Panels (d--f) show the numbers $N_F^M$
    of equivalent measurement results having probability $p_F^M(t)$.
  }  
\end{figure*}

We first analyze the time dependence of the measurement probabilities
$p_f^M(t)=|\braket{c^M_f|\psi(t)}|^2$. For that purpose,
we introduce a basis $(\ket{\Psi_\nu^{\rho,M}})$ of
the Hilbert space $\mathcal{H}$, each of whose
$2^N$ elements is an $N$--particle state which
is an eigenstate of $H$ with the energy $E_\nu^{\rho,M}$,
an eigenstate of $S_z$ with total spin projection $M$,
and transforms according to the irreducible representation $\rho$ of $D_{nh}$.
All its components
$\braket{c_f^M|\Psi_\nu^{\rho,M}}$ may be chosen real 
(owing to the fact that
all matrices $U_\phi$ of Sec.~\ref{sec:sym_space_spin} above
and all characters
of the irreducible representations of $D_{nh}$ are real \cite[\S 94]{landau3:BH1977}).
The probability amplitude $a_f^M(t)=\braket{c^M_f|\psi(t)}$ for the measurement
at time $t$ to yield the result $\ket{c_f^M}$, in the basis $\mathcal{C}$
with  total spin projection $M$, reads:
\begin{equation}
  \label{eq:timedep_probamplitude}
  a_f^M(t) 
  =\sum_\nu a_{f,\nu}^M\,  \exp\left(-i\omega_\nu^{M} \, t\right)
  \ ,
\end{equation}
where the coefficient
$a_{f,\nu}^M=\braket{c_f^M|\Psi_\nu^{\rho_1,M}}\braket{\Psi_\nu^{\rho_1,M}|\psi_0}$,
the frequency $\omega_\nu^M=E_\nu^{\rho_1,M}/\hbar$,
and
the sum over $\nu$
includes all eigenstates $\ket{\Psi_\nu^{\rho_1,M}}$ 
of $H$ in the subspace
$(\rho_1,M)$.
For all cases considered in this article, the components $\braket{c^M_f|\psi_0}$
are all real, though this is not a requirement. Then, the coefficients $a_{f,\nu}^M$
are real, and
the probability
$p_f^M(t)=|a_f^M(t)|^2$
is given by:
\begin{equation}
  \label{eq:timedep_prob}
  p_f^M(t)=\sum_\nu \left( a^M_{f,\nu} \right)^2
  +2\sum_{\nu<\nu'}
  a^M_{f,\nu} \, a^M_{f,\nu'} 
  \cos\left( \omega_{\nu,\nu'}^M\, t\right)\ ,
\end{equation}
where the transition frequency $\omega^M_{\nu,\nu'}=\omega^M_{\nu'}-\omega^M_{\nu}$.
Equation (\ref{eq:timedep_probamplitude}) shows that,
if the component $\ket{\psi_0^M}$ of the initial state with total spin
projection $M$ is non--zero,
the time dependence of
the probability amplitudes $\braket{c^M_f|\psi(t)}$   for
all $\binom{N}{n+M}$ possible measurement results $\ket{c_f^M}$
with total spin projection $M$
involve the same frequencies $\omega_\nu^M$, whose number
is the dimension $d=\dim(\rho_1,M)$. 
Hence, the  probabilities
$p_f^M(t)$ of Eq.~(\ref{eq:timedep_prob})
all share the same qualitative behavior,
piloted by $d$. If $d=1$
(which holds for all geometries of Fig.~\ref{fig:allgeoms} if $|M|=n$),
the probabilities $p_f^M(t)$ for measurement results with total spin
projection $M$
are constant. If $d=2$
(which holds for all geometries of Fig.~\ref{fig:allgeoms} if $|M|=n-1$),
the probabilities $p_f^M(t)$ all oscillate sinusoidally at the same
frequency $\omega_{1,2}$.
Finally, if $d\geq 3$,
the probabilities $p_f^M(t)$ undergo an aperiodic evolution
involving the same $d(d-1)/2$ frequencies $\omega_{\nu,\nu'}$,
with $1\leq \nu<\nu'\leq d$.

\subsection{ \label{sec:equiv_results}
  Equivalent and inequivalent measurement results
}
We now show that the spatial symmetries in $D_{nh}$
cause many measurement probabilities $p_f^M(t)$  to be equal at all times.
We consider two possible measurement results $\ket{c_f^M}$
and $\ket{c_{f'}^M}$ in the basis $\mathcal{C}$ with the same total spin
projection $M$. We call them `equivalent' if they correspond
to each other through a spatial symmetry, i.e.\ $\ket{c_{f'}^M}=U_\phi\ket{c_f^M}$
for some $\phi$ in $D_{nh}$, the unitary operator $U_\phi$ being defined
in Sec.~\ref{sec:sym_space_spin} above. The initial
$N$--particle state $\ket{\psi_0}$ transforms under the unit
representation $\rho_1$ of $D_{nh}$,
hence, so does the state $\ket{\psi(t)}=e^{-iHt/\hbar}\ket{\psi_0}$
just before the measurement. Therefore, the probability amplitude
$\braket{c_f^M|\psi(t)}=\braket{c_{f'}^M|U_\phi|\psi(t)}=\braket{c_{f'}^M|\psi(t)}$.
Thus, the probability amplitudes for equivalent measurement results are equal
at all times, and, hence, so are the corresponding measurement probabilities, 
$p_f^M(t)=p_{f'}^M(t)$.

Measurement results which do not correspond through
any symmetry operation are `inequivalent'. 
For a given value of  $M$, the number of
different measurement probabilities $p_F^M(t)$ is equal
to the number of inequivalent states $\ket{c_F^M}$ in the basis $\mathcal{C}$,
labeled with a capital `$F$'.
We prove in Appendix \ref{sec:different_probabilities},
using a known result from group theory,
that this number is equal to the dimension $\dim(\rho_1,M)$.
This is the dimension of the subspace
within which the component $\ket{\psi^{\rho_1,M}(t)}$ evolves.
It is also equal to the number of frequencies
entering Eq.~\eqref{eq:timedep_probamplitude}
(see Sec.~\ref{sec:qualitativebehavior} above).
Hence, counting the number of different probabilities $p_F^M(t)$
gives direct access to this number of frequencies, without resorting
to a Fourier transform.

For each of the different functions $p_F^M(t)$,
the number $N_F^M$
of equivalent measurement results $\ket{c_f^M}$ which share the same
measurement probability $p_F^M(t)$ is also entirely determined
by the spatial symmetries. It is the number of distinct
states $\ket{c_f^M}=U_\phi\ket{c_F^M}$, all in $\mathcal{C}$,
obtained by acting on $\ket{c^M_F}$ using all spatial symmetry
operators $U_\phi$ of Sec.~\ref{sec:sym_space_spin}.
The numbers $N_F^M$ satisfy
$\sum_{F=1}^{\dim(\rho_1,M)} N_F^M=\binom{N}{n+M}$.

\section{ \label{sec:threecases}
  Three cases within experimental reach}

\subsection{ \label{sec:XXZ_xi}
  The state $\ket{\xi}$ evolving under $H_\mathrm{XXZ}$}

We first illustrate the results of Sec.~\ref{sec:two_observable_consequences} above
on the case of the  initial state
$\ket{\psi_0}=\ket{\xi_{\vec{x}}}=\ket{\uparrow_1^x,\ldots,\uparrow_N^x}$,
which is the specific case of the states $\ket{\xi_{\vec{u}}}$, introduced in Sec.~\ref{sec:equiv_results},
for the spherical coordinates $(\theta=\pi/2,\phi=0)$.
We let it evolve under the XXZ Hamiltonian (i.e.\ $J\neq J_z$ in Eq.~\eqref{eq:XXZham}). The state $\ket{\xi_{\vec{x}}}$
is not an eigenstate of $H_\mathrm{XXZ}$, and the measurement probabilities $p_f^M(t)$
exhibit all three qualitative behaviors introduced in Sec.~\ref{sec:qualitativebehavior}
above, depending on the total projection $M$.

\emph{Constant probabilities for $M=\pm n$ ---}
For any value of $N=2n$, the probabilities $p_f^{M=\pm n}=1/2^N$ for the two 
measurement results $\ket{c_f^{M=\pm n}}=\ket{\uparrow_1^z,\ldots,\uparrow_1^z}$
and $\ket{\downarrow_1^z,\ldots,\downarrow_1^z}$ are constant,
owing to the subspaces
$(\rho_1,M=\pm n)$ having dimension 1.

\emph{Sinusoidal oscillations for $M=\pm (n-1)$ --- }
There are $N$ possible measurement results
in the basis $\mathcal{C}$ with total spin projection $M=n-1$. We label them
$\ket{c_f^{M=n-1}}=\ket{\uparrow_1^z\ldots\downarrow_f^z\ldots\uparrow_N^z}$
with $1\leq f\leq N$, where  the single 
$\ket{\downarrow^z}$ is located on site $f$:
for $1\leq f\leq n$, it is on one of the sites of the outer ring,
whereas for $n+1\leq f\leq N$, it is on the inner ring.
The subspace $(\rho_1,M=n-1)$ has dimension 2, being spanned by the two 
states
$\ket{e^{M=n-1}_\mathrm{outer}}=\sum_{f=1}^{n}\ket{c_f^{M=n-1}}/\sqrt{n}$
and
$\ket{e^{M=n-1}_\mathrm{inner}}=\sum_{f=n+1}^{N}\ket{c_f^{M=n-1}}/\sqrt{n}$.
The considerations of Sec.~\ref{sec:two_observable_consequences} then
yield the two following results. \textit{(A)} The $N$ measurement probabilities
$p_f^{M=n-1}(t)=|\braket{c_f^{M=n-1}|\psi(t)}|^2$ all oscillate sinusoidally
at the same frequency. \textit{(B)} There are two inequivalent measurement
results $\ket{c_{F}^{M=n-1}}$ with $F=1$ and $2$, which may be chosen as, say,
$\ket{c_{f=1}^{M=n-1}}$ and $\ket{c_{f=n+1}^{M=n-1}}$;
all 
$(\ket{c_f^{M=n-1}})_{1\leq f\leq n}$ share the same probability $p_1^{M=n-1}(t)$,
whereas all 
$(\ket{c_f^{M=n-1}})_{n+1\leq f\leq N}$ share the same probability
$p_{n+1}^{M=n-1}(t)$.
This behavior is a generalization of the  Rabi oscillation
\cite[Sec.~IV.C.3]{CDL1:Wiley2020}
to the case of $N=2n$ spins.
It affects the $N$ measurement results 
with total spin projection $M=-(n-1)$ in the same way.
It is a consequence of the double--ringed
nature of the considered trapping geometries.
It is not specific to the choice
of the initial state, and we shall encounter it again in Sec.~\ref{sec:Heisenberg_chi_collapse}
(see Fig.~\ref{fig:probs_timedep_chit6}a).

\emph{Aperiodic behavior for $1\leq |M|\leq n-2$ ---}
We consider a value of the total spin projection $M$
such that $1\leq |M|\leq n-2$. Then, 
for all considered geometries, 
$\dim(\rho_1,M)\geq 3$
(see Table~\ref{tab:groupth_results}).
Hence, the probabilities $p_f^M(t)=|\braket{c_f^M|\psi(t)}|^2$
exhibit an aperiodic dependence on $t$.
The special case of $M=0$ requires further analysis and is presented in
Sec.~\ref{sec:XXZ_twofoldspinrotation}
below.

\emph{Numerical results ---}
Panels (a) and (b) of Fig.~\ref{fig:probs_timedep_xit6} respectively
show the sinusoidal and aperiodic behaviors for the measurement
probabilities $p_f^M(t)$ with $M=2$ and $M=1$, obtained numerically
from the full $2^N\times 2^N$ Hamiltonian $H_\mathrm{XXZ}$
for $N=6$ particles, $J_z/J=-3$, and $\alpha=6$.
Panels (d) and (e) confirm that, in both cases,
the number of inequivalent measurement results $\ket{c_F^M}$
is equal to $\dim(\rho_1,M)$, and show the numbers $N^M_F$
of states $\ket{c^M_f}$ equivalent to each of them.
Our numerical results are 
in full agreement with
our predictions of the previous paragraphs based on symmetry arguments alone.

\begin{figure*}
  \includegraphics[width=\textwidth]
  {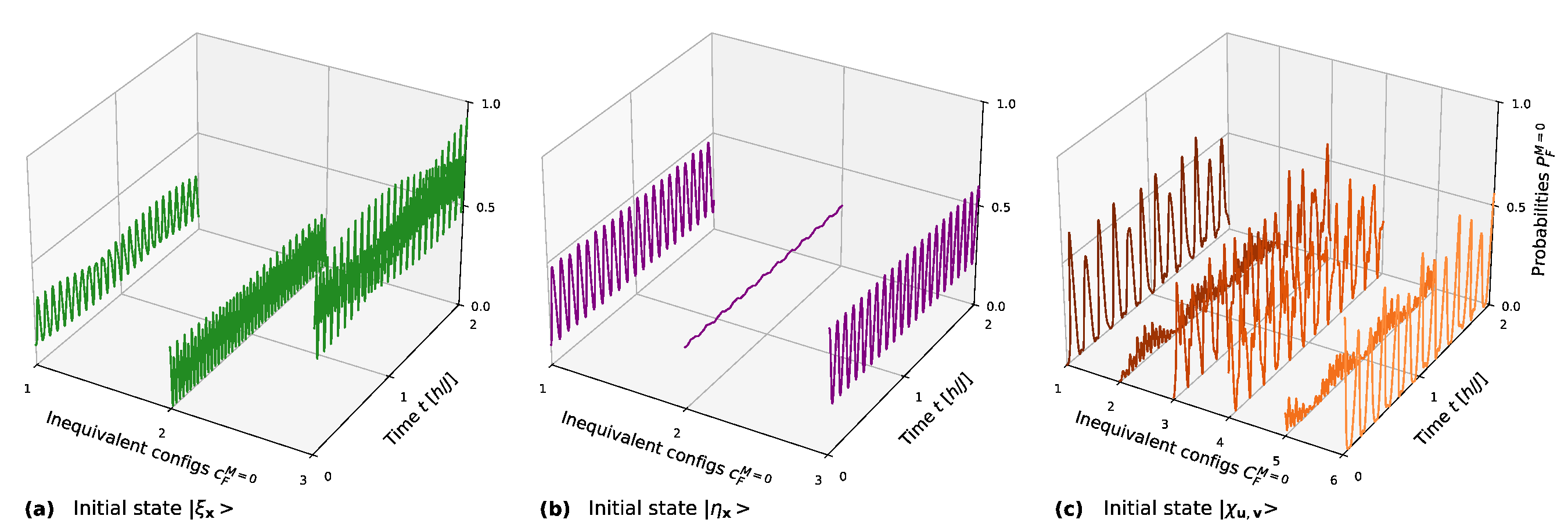}
  \includegraphics[width=\textwidth]
  {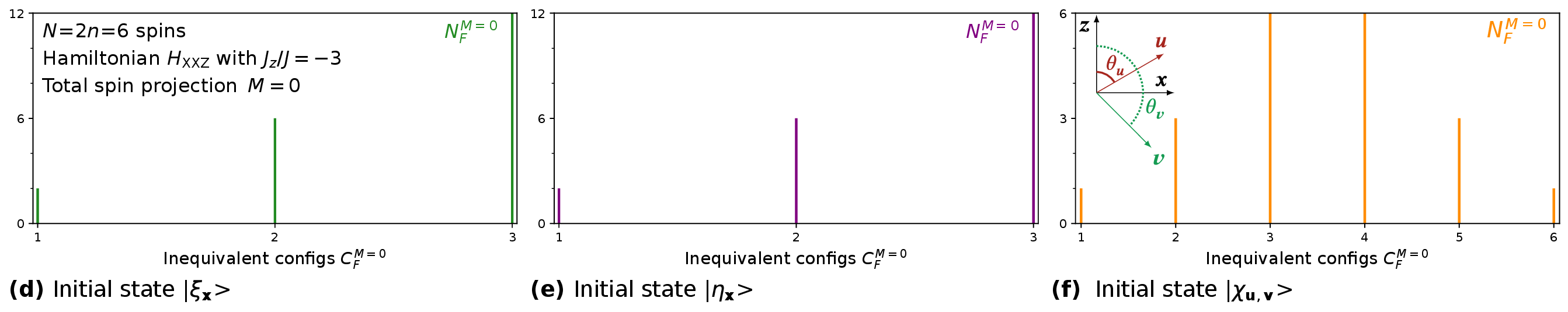}
  \caption{ \label{fig:M0Yspin}
    Panels (a,b,c):
    Time--dependent probabilities 
    for the
    inequivalent measurement results $\ket{c^{M=0}_F}$
    with total spin projection $M=0$,
     for the initial states
    $\ket{\psi_0}=\ket{\xi_{\vec{x}}}$ (a),
    $\ket{\eta_{\vec{x}}}$ (b), and
    $\ket{\chi_{\vec{u},\vec{v}}}$ (c).
    The unit vectors $\vec{u}$ and $\vec{v}$ on the Bloch sphere
    are chosen in the $(\vec{x},\vec{z})$ plane 
    with angles $\theta_{\vec{u}}=\pi/3$ and $\theta_{\vec{v}}=3\pi/4$
    [see inset to panel (f)].
    All states involve $N=6$ particles, and evolve under
    the Hamiltonian $H_\mathrm{XXZ}$ with $J_z/J=-3$ and $\alpha=6$.
    In each case,
    we represent the quantities
    $P_F^{M=0}(t)=p_F^{M=0}(t) \: N_F^{M=0}/\|\psi_0^{M=0}\|^2$
    (see Sec.~\ref{sec:XXZ_xi}).
    Panels (d,e,f) show the numbers $N_F^{M=0}$
    of equivalent measurement results having probability $p_F^{M=0}(t)$.
    [Panels (a,d) of this figure coincide with panels (c,f)
    of Fig.~\ref{fig:probs_timedep_xit6}.]
  }  
\end{figure*}

\emph{Convention used for representing the probabilities ---}
Panels (a--c) of Fig.~\ref{fig:probs_timedep_xit6} 
each focus on a given total spin projection $M$.
We show a single curve per inequivalent measurement result
$\ket{c_F^M}$,
and represent the quantities $P_F^M(t)=p_F^M(t)\: N_F^M/\|\psi_0^M\|^2$,
where $\|\psi_0^M\|^2=\braket{\psi_0^M|\psi_0^M}$, and
$\ket{\psi_0^M}$ is the component of $\ket{\psi_0}$ with total
spin projection $M$.
These are the total probabilities for
each set of equivalent measurement results,
rescaled such that $\sum_F P_F^M(t)=1$.
The same convention is used for 
Figs.~\ref{fig:M0Yspin} and \ref{fig:probs_timedep_chit6}
discussed below.

\subsection{ \label{sec:XXZ_twofoldspinrotation}
  XXZ Hamiltonian: two--fold rotation of the $N$ spins}

\begin{figure*}
  \includegraphics[width=\textwidth]
  {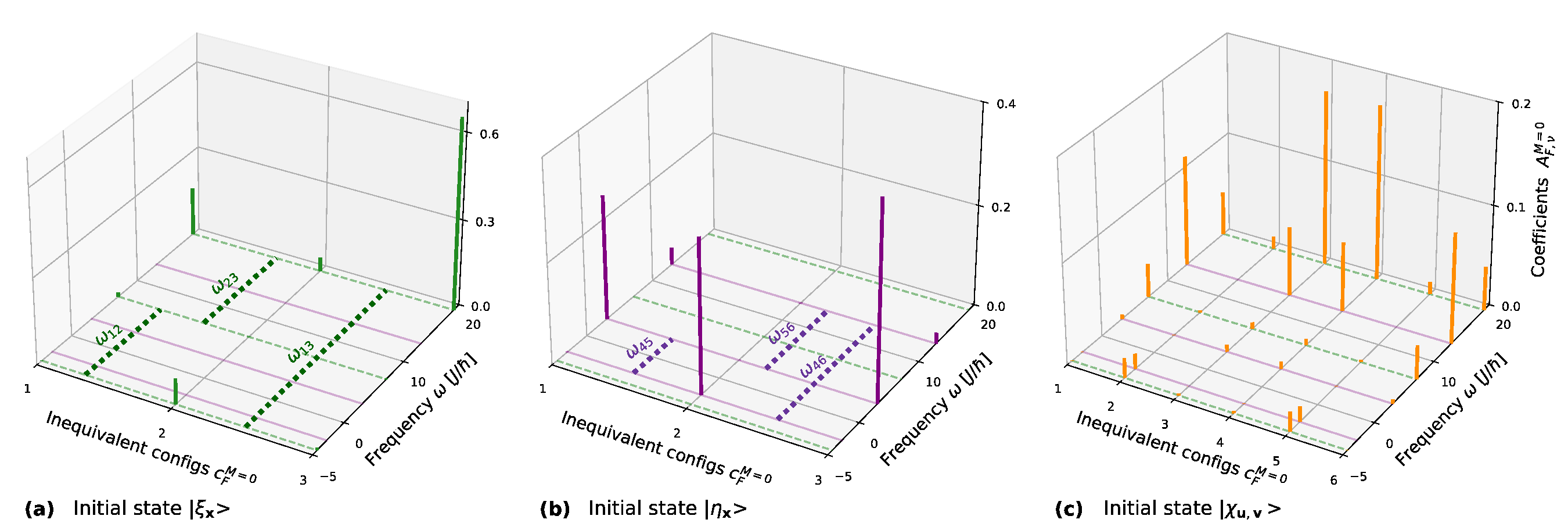}
  \includegraphics[width=\textwidth]
  {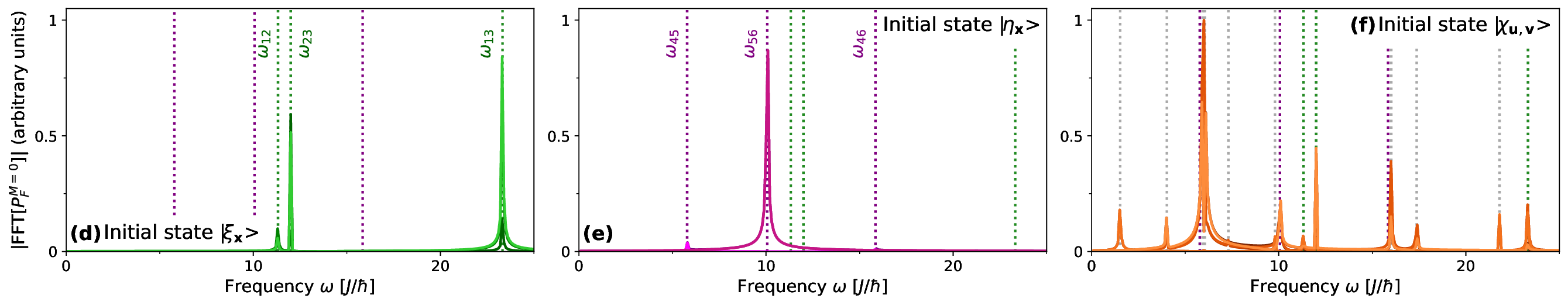}

  \caption{ \label{fig:MOYspin_Fourier}
    Panels (a,b,c): Fourier transforms of  the probability
    amplitudes $\braket{c_F^{M=0}|\psi(t)}$ obtained for the initial states
    and Hamiltonians of Fig.~\ref{fig:M0Yspin}.
    We represent the
    quantities
    $A_{F,\nu}^{M=0}=N_F^{M=0} \: (a_{F,\nu}^{M=0}/\|\psi_0^{M=0}\|)^2$,
    introduced in Sec.~\ref{sec:oddn},
    as a function of the frequencies $\omega_\nu^{M=0}$.
    The energies $\hbar\omega^{M=0}_\nu$ of the eigenstates of $H_\mathrm{XXZ}$
    which are odd and even under the operator $Y_\mathrm{spin}$
    are respectively shown 
    as thin green and purple  horizontal lines on the bases of the three plots.
    The thick dashed  lines on the bases of (a) and (b) show the transition
    frequencies $\omega_{\nu,\nu'}$
    respectively allowed for $\ket{\xi_{\vec{x}}}$ and $\ket{\eta_{\vec{x}}}$.
    [The scales
    along the vertical axes of the three panels (a,b,c) are different.]
    Panels (d,e,f): absolute values of the Fast Fourier Transforms (FFT) of the corresponding
    probabilities $P_F^{M=0}(t)$
    for $0\leq t\leq t_\mathrm{max}=10\, h/J$, all superimposed.
    The transition frequencies $\omega_{\nu,\nu'}$ allowed for
    $\ket{\xi_{\vec{x}}}$ and $\ket{\eta_{\vec{x}}}$ 
    are shown as the vertical green and purple lines. All other
    transition frequencies are shown in gray on panel (f).
  }
\end{figure*}

The conservation laws 
stated in Sec.~\ref{sec:conservationlaws_measurement}
account for all spatial symmetries in $D_{nh}$ and spin rotations
about the axis $\vec{z}$ through arbitrary angles.
However,
for some choices of the Hamiltonian and initial state,
additional spin symmetries 
come into play.
Then, the numbers of frequencies and inequivalent measurement results
obtained in
Sec.~\ref{sec:two_observable_consequences}
are overestimates, which may be refined.
The case of the initial state $\ket{\xi_{\vec{x}}}$ evolving under
the Heisenberg Hamiltonian, discussed in Sec.~\ref{sec:initstates} above,
is a simple example. In this case, the conservation of $\vec{S}^2$
entails that the component
$\ket{\psi^{\rho_1,M}(t)}$ of the $N$--particle state $\ket{\psi(t)}$
with total spin projection $M$
is proportional to the single eigenstate of $H_H$
in the  subspace $(\rho_1,M)$ with maximal total spin modulus $S=n$:
thus, for a given $M$, all
probabilities $p_f^M=|\braket{c_f^M|\psi(t)}|^2$ are constant and equal.

We now consider the spin rotation $g=C^{\vec{e}}_\theta$
through the angle $\theta$ about an arbitrary axis $\vec{e}$
in the horizontal plane $(xy)$. For the Heisenberg Hamiltonian,
$C^{\vec{e}}_\theta$ is an element of
the spin symmetry group $G^\mathrm{spin}$ of Sec.~\ref{sec:sym_space_spin}
above
for any angle $\theta$,
leading to the conservation of the operator $\vec{e}\cdot\vec{S}$
(this actually holds for any direction $\vec{e}$, which may be chosen
instead of $\vec{z}$
as the quantization axis).
By contrast, for the XXZ Hamiltonian,
$C^{\vec{e}}_\theta$ is in $G^\mathrm{spin}$
only for $\theta=0$ or $\pi$.
In the remainder of the present section \ref{sec:XXZ_twofoldspinrotation},
we focus on the XXZ case 
and identify the parity conservation law corresponding to these spin rotations.
Then, we demonstrate its role
by comparing the measurement probabilities obtained from
three different initial states, discussing both their time dependence
and their Fourier transform.

\subsubsection{ \label{sec:conservationlaw_M=0_Yspin}
  Parity under $Y_\mathrm{spin}$ for quantum states with $M=0$}
We consider
the spin rotation $g=C_{\pi}^{\vec{y}}$ through angle $\pi$
about the axis $\vec{y}$. 
The operator $Y_\mathrm{spin}$,
which acts on the Hilbert space $\mathcal{H}$
and represents $g$,
reads (see Sec.~\ref{sec:sym_space_spin} above):
\begin{equation} \label{eq:Yspin}
  Y_{\mathrm{spin}}= U_{g}=(-i\sigma_1^y)\ldots (-i\sigma_N^y)
  \ .
\end{equation}
The operator $Y_{\mathrm{spin}}^2=1$, because  the system is comprised of an
even number $N=2n$ of spins--$1/2$ \cite[\S 99]{landau3:BH1977}.
Thus, $Y_\mathrm{spin}$ represents a two--fold rotation of the $N$ spins.

We focus  on the the subspace $(\rho_1,M=0)$ defined in
Sec.~\ref{sec:two_observable_consequences}, which is invariant under $Y_\mathrm{spin}$.
It is the direct sum of two subspaces,
$
(\rho_1,M=0)=
(\rho_1,M=0,\mathrm{even/}Y_\mathrm{spin})
\oplus
(\rho_1,M=0,\mathrm{odd/}Y_\mathrm{spin})
$,
respectively comprised of the eigenstates of $Y_\mathrm{spin}$
with eigenvalue $+1$
(states which are even under  $Y_\mathrm{spin}$,
denoted $\mathrm{even}/Y_\mathrm{spin}$)
and $-1$
(states which are odd under $Y_\mathrm{spin}$,
denoted $\mathrm{odd}/Y_\mathrm{spin}$).
The dimensions of these two subspaces
are given in Table \ref{tab:groupth_results} for all four
geometries of Fig.~\ref{fig:allgeoms}.
We prove in Appendix \ref{sec:permrep:twofoldspinrotation}
that all spin rotations $C_{\pi}^{\vec{e}}$ through angle $\pi$ about an arbitrary
axis $\vec{e}$ in the $(xy)$ plane act on states with total spin projection
$M=0$ as the operator $Y_\mathrm{spin}$ of Eq.~\eqref{eq:Yspin}
and, hence, lead to the same definition of parity.

The parity with respect to $Y_\mathrm{spin}$ is conserved
during the evolution described by the Schr\"odinger equation.
However, the possible measurement results
$\ket{c_f^{M=0}}$ in the basis $\mathcal{C}$
each have both even and odd components under
$Y_\mathrm{spin}$, respectively given by
$(1\pm Y_\mathrm{spin})\ket{c_f^{M=0}}/\sqrt{2}$.
Thus,
in general, the probability amplitudes
$\braket{c_f^{M=0}|\psi(t)}$ result from the interference between
the components of $\ket{\psi(t)}$ along the even and odd subspaces.
Extending the idea  introduced in Sec.~\ref{sec:initstates} above,
we avoid this interference by choosing initial states $\ket{\psi_0}$
whose components $\ket{\psi_0^{M=0}}$ with total spin projection $M=0$
are eigenstates of $Y_\mathrm{spin}$. We show in Appendix~\ref{sec:initialstates_parityYspin}
that,
among the states $\ket{\chi_{\vec{u},\vec{v}}}$
of the form of Eq.~\eqref{eq:def_chi_uv}, only
two families of states satisfy this property. Both families are defined in terms of
one arbitrary unit vector $\vec{u}$ representing a direction on the Bloch sphere:
\textit{(i)} the states
$\ket{\xi_{\vec{u}}}=\ket{\chi_{\vec{u},\vec{u}}}$ introduced in Sec.~\ref{sec:initstates}, 
and \textit{(ii)} the states
$\ket{\eta_{\vec{u}}}=\ket{\chi_{\vec{u},\vec{u}'}}$, where the unit vector $\vec{u}'$
is the image of $\vec{u}$ under the rotation through
angle $\pi$ about the axis $\vec{z}$.
The components $\ket{\xi^{M=0}_{\vec{u}}}$ and $\ket{\eta^{M=0}_{\vec{u}}}$
are eigenstates of $Y_\mathrm{spin}$ corresponding to the eigenvalues
$(-1)^n$ and $+1$, respectively.

If the initial states $\ket{\psi_0}=\ket{\xi_{\vec{u}}}$ or $\ket{\eta_{\vec{u}}}$
are chosen in the protocol ($\mathcal{P}$), the consequences (A) and (B)
of symmetry described in Sec.~\ref{sec:two_observable_consequences} are strengthened
as follows. (A) 
The number of different frequencies entering the probability amplitudes
$\braket{c_f^{M=0}|\psi(t)}$ with total spin projection
$M=0$, given by Eq.~\eqref{eq:timedep_probamplitude},
and (B) the number of inequivalent measurement results $\ket{c_f^{M=0}}$,
are both equal to the dimension $d=\dim(\rho_1,M=0,\mathrm{even}/Y_\mathrm{spin})$.
For odd values of $n$, this dimension is also equal to $\dim(\rho_1,M=0,\mathrm{odd}/Y_\mathrm{spin})$
(see Table~\ref{tab:groupth_results}).
We prove these properties in Appendix \ref{sec:different_probabilities}.
We
have checked them numerically in all four geometries of Fig.~\ref{fig:allgeoms}.
In particular,
our numerical results for the initial state $\ket{\xi_{\vec{x}}}$ in the geometry with
$N=2n=6$,
and with the same parameters as in Sec.~\ref{sec:XXZ_xi},
are shown on panels (c) and (f) of Fig.~\ref{fig:probs_timedep_xit6}:
they confirm the presence of $\dim(\rho_1,M=0,\mathrm{even}/Y_\mathrm{spin})=3$
[rather than $\dim(\rho_1,M=0)=6$] inequivalent measurement results.

To summarize, three different cases are accessible using initial states
$\ket{\psi_0}=\ket{\chi_{\vec{u},\vec{v}}}$ of the form of Eq.~\eqref{eq:def_chi_uv}:
\textit{(i)} $n$ even, $\ket{\psi_0^{M=0}}$ even under $Y_\mathrm{spin}$;
\textit{(ii)} $n$ odd, $\ket{\psi_0^{M=0}}$ even under $Y_\mathrm{spin}$;
\textit{(iii)} $n$ odd, $\ket{\psi_0^{M=0}}$ odd under $Y_\mathrm{spin}$.
In all cases, (A) the number of frequencies entering the probability
amplitudes $\braket{c_f^{M=0}|\psi(t)}$ and (B) the number of inequivalent
measurement results are both equal to the dimension of the subspace
within which the component $\ket{\psi^{M=0}(t)}$ evolves, as 
in Sec.~\ref{sec:equiv_results}. The additional conservation law
of parity under $Y_\mathrm{spin}$
further constrains the dimension of this subspace, which is now
$(\rho_1,M=0,\text{even}/Y_\mathrm{spin})$ or $(\rho_1,M=0,\text{odd}/Y_\mathrm{spin})$.

\subsubsection{\label{sec:oddn}%
  Case of odd $n$: three different initial states evolving under $H_\mathrm{XXZ}$}

In this section, we identify observable consequences of the
conservation of parity under $Y_\mathrm{spin}$. In particular,
we show how to demonstrate the fact
that initial states with total spin projection $M=0$
which are even or odd under $Y_\mathrm{spin}$
give rive to quantum dynamics occurring within
the different subspaces
$(\rho_1,M=0,\mathrm{even}/Y_\mathrm{spin})$ or
$(\rho_1,M=0,\mathrm{odd}/Y_\mathrm{spin})$, respectively.
We consider the initial states
$\ket{\xi_{\vec{u}}}$ and $\ket{\eta_{\vec{u}}}$,
which are the only $N$--particle states of the form of Eq.~\eqref{eq:def_chi_uv}
whose $M=0$ component is an eigenstate of $Y_\mathrm{spin}$
(see Sec.~\ref{sec:conservationlaw_M=0_Yspin} above).
The direction $\vec{u}$ may be chosen arbitrarily on the Bloch sphere.
For even values of $n$, the components
$\ket{\xi^{M=0}_{\vec{u}}}$ and $\ket{\eta^{M=0}_{\vec{u}}}$
are both even under $Y_\mathrm{spin}$, so that odd states 
are inaccessible with the considered
initial states, and the dynamics of even and odd states may not be compared.
By contrast, for odd values of $n$, 
$\ket{\xi^{M=0}_{\vec{u}}}$ and $\ket{\eta^{M=0}_{\vec{u}}}$
are respectively odd and even under $Y_\mathrm{spin}$.
Therefore, we focus on the case of odd $n$ and compare the
measurement probabilities obtained from these two states.

\emph{Time dependence ---}
In Fig.~\ref{fig:M0Yspin},
we compare the probabilities $p_f^{M=0}(t)$
for the measurement results with total spin projection $M=0$,
for the geometry involving $N=2n=6$ spins,
and for three
different initial states $\ket{\psi_0}=$
$\ket{\xi_{\vec{x}}}$, $\ket{\eta_{\vec{x}}}$, and
$\ket{\chi_{\vec{u},\vec{v}}}$.
For the state $\ket{\chi_{\vec{u},\vec{v}}}$,
the directions $\vec{u}$ and $\vec{v}$ of the Bloch
sphere are chosen in the $(x,z)$ plane, the angles
$\theta_{\vec{u}}=\pi/3$ and $\theta_{\vec{v}}=3\pi/4$
being defined in the inset to Fig.~\ref{fig:M0Yspin}f.
These three initial states evolve under the same Hamiltonian $H_\mathrm{XXZ}$.
The parameters $J$, $J_z$ and $\alpha$ entering Eq.~\eqref{eq:XXZham}
are the same as in Fig.~\ref{fig:probs_timedep_xit6},
so that panels (a,d) of Fig.~\ref{fig:M0Yspin} coincide with
panels (c,f) of Fig.~\ref{fig:probs_timedep_xit6}.
The components $\ket{\xi^{M=0}_{\vec{x}}}$ and $\ket{\eta^{M=0}_{\vec{x}}}$
each give rise to $3$ inequivalent measurement results
$\ket{c_F^{M=0}}$ 
(see panels d, e of Fig.~\ref{fig:M0Yspin}), in full agreement
with our prediction of Sec.~\ref{sec:conservationlaw_M=0_Yspin}.
By contrast, the component $\ket{\chi^{M=0}_{\vec{u},\vec{v}}}$,
which is not an eigenstate of $Y_\mathrm{spin}$, yields
six inequivalent measurement results (panels c and f), in accordance
with the result of Sec.~\ref{sec:equiv_results}.

\emph{Fourier transform ---}
We now identify an observable signature of the fact that initial states
whose $M=0$ components, $\ket{\psi^{M=0}_0}$, are even or odd under $Y_\mathrm{spin}$,
yield quantum evolutions for $\ket{\psi^{M=0}(t)}$ occurring within
different subspaces. 
We introduce the Fourier transform
$\widetilde f(\omega)=\int_{-\infty}^{\infty}dt\, e^{i\omega t}f(t)/(2\pi)$
of any function of time $f(t)$.
The Fourier transform $\widetilde a^M_f(\omega)$
of the probability amplitude $a_f^M(t)$ 
of Eq.~\eqref{eq:timedep_probamplitude}
reads, for $M=0$,
\begin{equation}
  \label{eq:probamplitude_Fourier}
  \widetilde{a}_f^{M=0}(\omega)=\sum_\nu a_{f,\nu}^{M=0} \:\delta(\omega-\omega^{M=0}_\nu) 
  \ .
\end{equation}
The frequencies $\omega^{M=0}_\nu=E^{\rho_1,M=0}_\nu/\hbar$
entering Eq.~\eqref{eq:probamplitude_Fourier}
are determined by
the eigenvalues of $H_\mathrm{XXZ}$ for the eigenstates
in the subspace $(\rho_1,M=0)$, which
do not depend on the initial state $\ket{\psi_0}$.
By contrast, the coefficients
$a_{f,\nu}^{M=0}$ are proportional to $\braket{\Psi_\nu^{\rho_1,M=0}|\psi_0}$
(see Sec.~\ref{sec:qualitativebehavior})
and, hence, do depend on $\ket{\psi_0}$.
The eigenstates $\ket{\Psi^{\rho_1,M=0}_\nu}$  may each be chosen to be
either even or odd under $Y_\mathrm{spin}$.
Thus, if $\ket{\psi_0^{M=0}}$ is even (resp.\ odd) 
under $Y_\mathrm{spin}$, 
only even (resp.\ odd) eigenstates
take part in
Eq.~\eqref{eq:probamplitude_Fourier}, and
$\widetilde{a}_f^{M=0}(\omega)$ is non--zero
only for the corresponding subset of frequencies $\omega^{M=0}_\nu$.
By contrast, if $\ket{\psi_0^{M=0}}$ is not an eigenstate of $Y_\mathrm{spin}$,
all frequencies $\omega^{M=0}_\nu$  take part in the sum.

The Fourier transform $\widetilde p_f^{M=0}(\omega)$
of the probability  $p_f^{M=0}(t)$ given by Eq.~\eqref{eq:timedep_prob},
reflects the parity of $\ket{\psi_0^{M=0}}$ under $Y_\mathrm{spin}$
similarly.
If $\ket{\psi_0^{M=0}}$ is even (resp.\ odd),  
the Fourier transform $\widetilde p_f^{M=0}(\omega)$
only involves transition frequencies $\omega_{\nu,\nu'}$ corresponding
to pairs of eigenstates $\ket{\Psi_{\nu}^{\rho_1,M=0}}$ and $\ket{\Psi_{\nu'}^{\rho_1,M=0}}$
which are both even (resp.\ odd) under $Y_\mathrm{spin}$.
By contrast,
if $\ket{\psi_0^{M=0}}$ is not an eigenstate of $Y_\mathrm{spin}$,
the transition frequencies corresponding to all pairs of eigenstates
in the $(\rho_1,M=0)$ subspace, regardless of their parity,
may enter $\widetilde p_f^{M=0}(\omega)$.

\begin{figure*}
  \includegraphics[width=\linewidth]
  {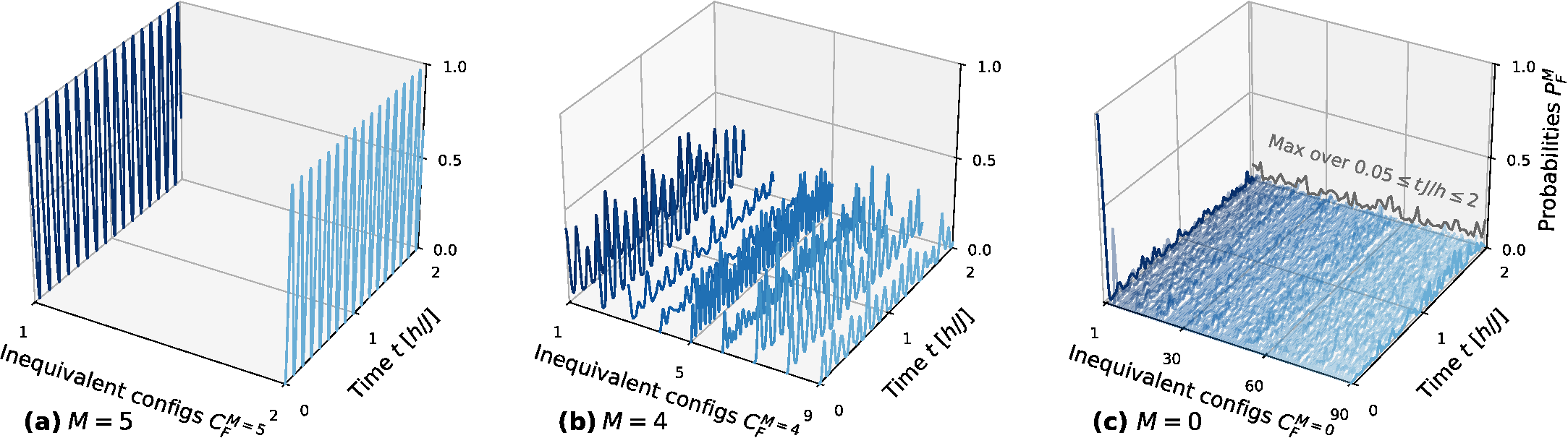}
  \includegraphics[width=\linewidth]
  {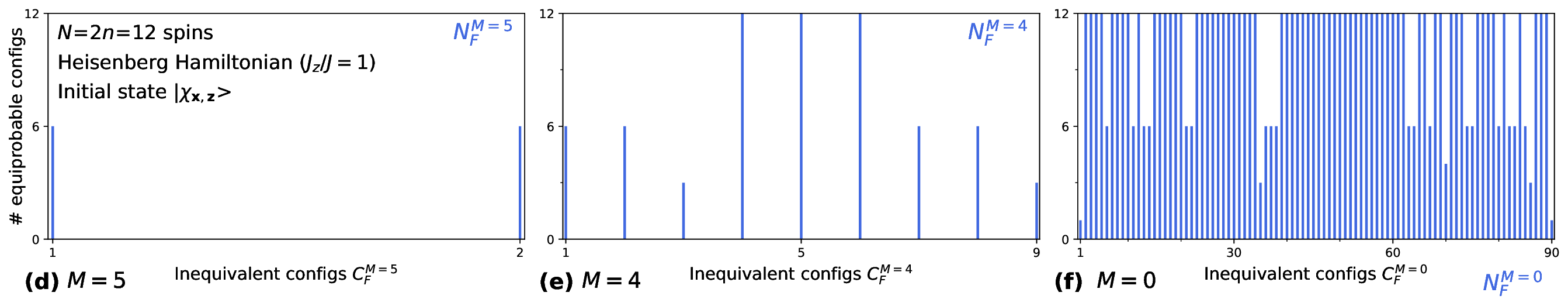}
  \caption{ \label{fig:probs_timedep_chit6}
    Time--dependent probabilities 
    for the inequivalent measurement results $\ket{c^M_F}$
    with total spin projection $M$,
    for the 
    initial state $\ket{\psi_0}=\ket{\chi_{\vec{x},\vec{z}}}$
    involving $N=12$ particles,
    evolving under the Heisenberg Hamiltonian 
    with $\alpha=6$,
    for $M=5$ (a), $4$ (b), and $0$ (c).
    For each $M$, we represent the ratios
    $P_F^M(t)=p_F^{M}(t) \: N_F^{M}/\|\psi_0^{M}\|^2$.
    The gray curve on the back face of panel (c)
    shows the maximum value of each $P_F^{M=0}(t)$
    over the time interval $0.05\leq t\, J/h\leq 2$.
    Panels (d--f) show the numbers $N_F^M$ of equivalent measurement results having
    probability $p_F^M(t)$.
  }
\end{figure*}

These predictions are fully confirmed by our numerical results
illustrated in Fig.~\ref{fig:MOYspin_Fourier}.
Its panels (a--c) 
show the coefficients $a_{f,\nu}^{M=0}$
entering Eq.~\eqref{eq:probamplitude_Fourier} above
as a function of the frequencies $\omega_\nu^{M=0}$,
for the three initial
states of Fig.~\ref{fig:M0Yspin}, all evolving under the Hamiltonian $H_\mathrm{XXZ}$
with the same parameters.
In each case, we show a single set of coefficients 
per inequivalent result $\ket{c_F^{M=0}}$,
and represent the quantities
$A_{F,\nu}^{M=0}=N_F^{M=0}\: (a_{F,\nu}^{M=0})^2/\|\psi_0^{M=0}\|^2$,
whose sum over $F$ and $\nu$ is 1.
The state $\ket{\xi_{\vec{x}}}$, whose $M=0$ component is odd under $Y_\mathrm{spin}$,
has non--zero coefficients $a^{M=0}_{F,\nu}$ only for
the three frequencies  $\omega^{M=0}_\nu=E_\nu^{\rho_1,M=0}/\hbar$
corresponding to eigenstates of $H_\mathrm{XXZ}$ which are odd under $Y_\mathrm{spin}$,
shown as the thin solid green lines on the base of each plot.
The state $\ket{\eta_{\vec{x}}}$, whose $M=0$ component is even under $Y_\mathrm{spin}$,
has non--zero coefficients $a^{M=0}_{F,\nu}$ only for
the three frequencies $\omega^{M=0}_\nu$ corresponding
to even eigenstates, shown as the thin dashed purple lines.
By contrast, the state $\ket{\chi_{\vec{u},\vec{v}}}$ has non--zero coefficients $a^{M=0}_{F,\nu}$
for all six eigenstates.
Panels (d--f) represent the Fast Fourier Transform (FFT)
of the probabilities $P_F^{M=0}(t)$
over the time interval $0\leq t\leq 10h/J$,
for the inequivalent results $\ket{c_F^{M=0}}$, all superimposed. Panel (d)
exhibits three peaks, corresponding to the three transition frequencies
$\omega_{12}$, $\omega_{23}$, $\omega_{31}$ between odd states,
shown by the vertical dashed green lines and identified on
the base of panel (a). Panel (e) shows one dominant peak for $\omega_{56}$
and two smaller ones for $\omega_{45}$ and $\omega_{46}$,
corresponding to the three transition frequencies between even states,
shown by the vertical dashed purple lines and
identified on the base of panel (b). 
Finally, panel (c) shows numerous peaks, some of which occur for transition frequencies
involving even or odd states, and others for transition frequencies involving
an even state and an odd one, shown by the vertical dashed gray lines.

The peaks of panels (a) occur
for frequencies $(\omega_{12},\omega_{23},\omega_{31})$  which are all different
from those of panel (b), namely, $(\omega_{45},\omega_{56},\omega_{64})$.
This provides the sought signature of the quantum evolution within different subspaces
for initial states whose component $\ket{\psi_0^{M=0}}$ is odd or even under $Y_\mathrm{spin}$.
Panel (c) further illustrates that, if $\ket{\psi_0^{M=0}}$ is not an eigenstate of
$Y_\mathrm{spin}$, the measurement performed at time $t$ causes interference between the
even and odd components of $\ket{\psi^{M=0}(t)}$.
All these predictions may readily be tested in experiments by taking the FFT of the
time--dependent measurement probabilities $p_F^M(t)$.

\subsection{ \label{sec:Heisenberg_chi_collapse}
  For larger spin numbers $N$: collapse of the initial state}

\begin{figure} 
  \includegraphics[width=\linewidth]
  {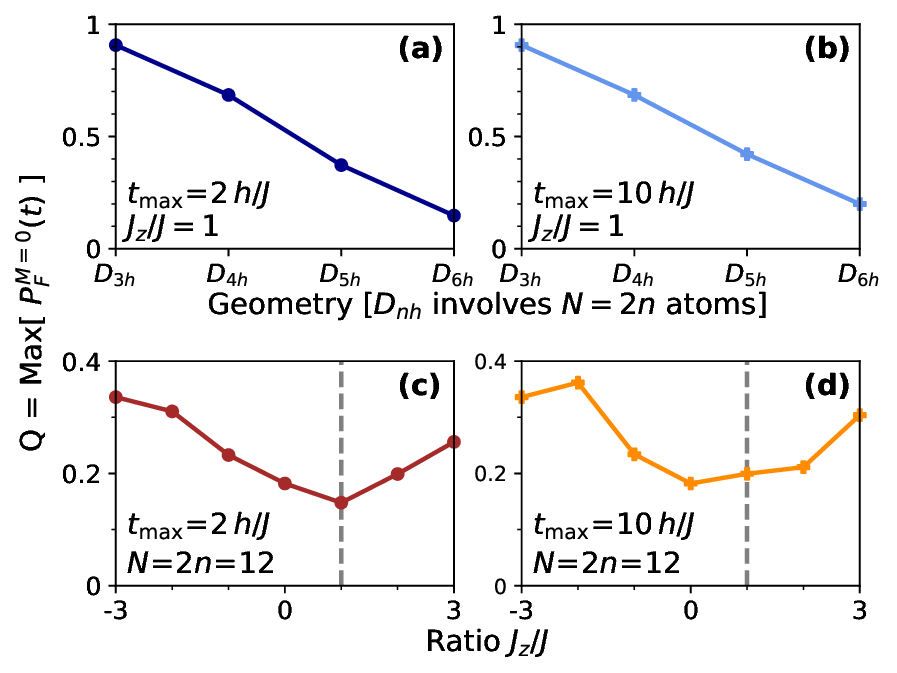}
  \caption{ \label{fig:compare_geometries_JzovJ}
    Collapse of the component of the initial state
    $\ket{\psi_0}=\ket{\chi_{\vec{x},\vec{z}}}$
    with total spin projection $M=0$. On each panel,
    we represent the maximum $Q$ of the quantities
    $P^{M=0}_F(t)=p^{M=0}_F(t)\: N_F^{M=0}/\|\psi_0^{M=0}\|^2$
    over all
    inequivalent measurement results with $M=0$
    and times $t$ such that
    $0.05\,h/J\leq t\leq t_\mathrm{max}$.
    On panels (a,b),
    we compare the four systems of
    Fig.~\ref{fig:allgeoms} evolving under the Heisenberg
    Hamiltonian ($J_z/J=1$).
    On panels (c,d), we consider the system comprised of $N=2n=12$ spins, 
    and vary the ratio  $J_z/J$ from $-3$ to $3$
    (the dashed
    gray line shows the Heisenberg value $J_z/J=1$).
    The maximum time $t_\mathrm{max}=2\,h/J$ for panels (a,c)
    and $10\,h/J$ for panels (b,d).
  }
\end{figure}

\subsubsection{ \label{sec:Heisenberg_chi_collapse_M>0}
  Evolution under the Heisenberg Hamiltonian}

In this section, we choose the $N$--particle initial state
$\ket{\psi_0}=\ket{\chi_{\vec{x},\vec{z}}}$,
given by  Eq.~\eqref{eq:def_chi_uv} with
$\vec{u}=\vec{x}$ and $\vec{v}=\vec{z}$,
and let the system evolve under the Heisenberg Hamiltonian.

For total spin projections $M>0$, the resulting
measurement probabilities $p_F^M(t)$ behave in a very
similar way as those 
obtained from the initial state $\ket{\xi_{\vec{x}}}$ evolving under the
XXZ Hamiltonian (see Sec.~\ref{sec:XXZ_xi} above).
Specifically, the single probability with maximal $M=n$ is constant;
the $N$ probabilities with $M=n-1$, among which two are inequivalent,
oscillate sinusoidally; the probabilities $p_F^M(t)$
with $1\leq M\leq n-2$ undergo an aperiodic evolution.
These predictions are confirmed by our numerical results for $N=12$ spins,
illustrated in Fig.~\ref{fig:probs_timedep_chit6} for $N=2n=12$ spins.
Its panels (a) and (b) show the sinusoidal and aperiodic behaviors
expected for $M=5$ and $4$, respectively, and the corresponding panels (d) and (e)
show the numbers of equivalent measurement results $N_F^M$. 
They are directly comparable to panels (a,b) and (d,e)
of Fig.~\ref{fig:probs_timedep_xit6} above.

Despite the choice of the Hamiltonian $H_\mathrm{H}$, these results
exhibit no straightforward signature of the conservation of the total spin modulus $S$.
This is because 
the component
$\ket{\chi^{M}_{\vec{x},\vec{z}}}$ of the initial state with total spin projection
$M$ is not an eigenstate of the squared total spin operator $\vec{S}^2$
(except for $M=n$).
In particular, the sinusoidal regime of $M=n-1$ involves two eigenstates
of $H_\mathrm{H}$ with $S=n$ and $S=n-1$, respectively.
These states are not coupled during the evolution described by the Schrödinger equation.
However, the measurement at time $t$ causes them to interfere,
because the configurations $\ket{c_f^{M=n-1}}$
have non--zero components with both $S=n$ and $S=n-1$.

\subsubsection{ \label{sec:Heisenberg_chi_collapse_M=0}
  Collapse of the $M=0$ component of the initial state}

We retain the  initial state $\ket{\chi_{\vec{x},\vec{z}}}$ evolving under the
Heisenberg Hamiltonian, 
and turn to the measurement probabilities $p_F^{M=0}(t)$ with total spin projection $M=0$.
They exhibit a specific qualitative behavior,
dictated by the two following properties.
Firstly,
the component $\ket{\chi_{\vec{x},\vec{z}}^{M=0}}$ of the initial
state  reads:
\begin{equation} \label{eq:chixz_M0}
  \ket{\chi_{\vec{x},\vec{z}}^{M=0}}=
  \frac{1}{2^{n/2}}
  \ket{\downarrow_1^z,\ldots\downarrow_n^z;\uparrow_{n+1}^z,\ldots,\uparrow_N^z}
  =
  \frac{1}{2^{n/2}}\ket{c_1^{M=0}}
  \ ,
\end{equation}
the configurations $\ket{c_F^M}$ being numbered as in Appendix \ref{sec:config_numbering}.
Equation~\eqref{eq:chixz_M0} shows that $\ket{\chi_{\vec{x},\vec{z}}^{M=0}}$ is proportional
to the single configuration $\ket{c_1^{M=0}}$. 
Secondly, for larger values of $N$,
the dimension of the subspace $(\rho_1,M=0)$ increases.
For example, this dimension is $48+42=90$ for $N=12$ (see Table~\ref{tab:groupth_results}).

The combination of these two properties yields
the collapse of the initial component $\ket{\chi^{M=0}_{\vec{x},\vec{z}}}$,
illustrated in Fig.~\ref{fig:probs_timedep_chit6}c for $N=2n=12$.
We now discuss this specific case.
Initially, the only configuration $\ket{c_F^{M=0}}$ with non--zero probability
is $\ket{c_1^{M=0}}$.
Thus, the quantity $P_{1}^{M=0}(t)$,
introduced in Sec.~\ref{sec:XXZ_xi} and
represented in Fig.~\ref{fig:probs_timedep_chit6}c,
satisfies $P_{1}^{M=0}(0)=1$, and all other  $P_{F}^{M=0}(0)=0$.
The value of $P_{1}^{M=0}(t)$ strongly decreases over a very short time $t$,
and after a transient regime whose duration is of the order of $0.05h/J$,
the quantities $P_{F}^{M=0}(t)$ for all inequivalent measurement results
$\ket{c_F^{M=0}}$ remain $<0.15$ for all times up to $t_\mathrm{max}=2h/J$,
as shown by the gray curve on the back face of Fig.~\ref{fig:probs_timedep_chit6}.
The quantities $P_{F}^{M=0}(t)$ corresponding to different numbers
of equivalent probabilities $N_F^{M=0}$ (shown in Fig.~\ref{fig:probs_timedep_chit6}f)
have comparable magnitudes.
In particular, nine quantities  $P_F^{M=0}(t)$ exceed $0.1$ at least once
over the time interval $0.05h/J<t<2h/J$, with
$N_F^{M=0}$  equal to $1$, $6$, or $12$.

Our numerical results exhibit no revival of the initial state for longer durations
up to $t_\mathrm{max}=10h/J$.

\subsubsection{Comparison of various system sizes and Hamiltonians}

Finally, starting from the same initial state $\ket{\chi_{\vec{x},\vec{z}}}$
as in Sec.~\ref{sec:Heisenberg_chi_collapse_M=0}
above, we seek to optimize the observation of the collapse by
varying the number of spins $N=2n$ or the ratio $J_z/J$ entering
the Hamiltonian of Eq.~\eqref{eq:XXZham}. We characterize the quality of the
collapse by the maximum $Q=\max[P_F^{M=0}(t)]$, taken over all inequivalent
measurement results $\ket{c_F^{M=0}}$ with total spin projection $M=0$,
and all times $t$ such that $0.05\, h/J\leq t\leq t_\mathrm{max}$,
where $t_\mathrm{max}=2\, h/J$ or $10\, h/J$. Lower values of $Q$
signal a higher quality for the collapse.

Our results are summarized in Fig.~\ref{fig:compare_geometries_JzovJ}.
We first assume that the system evolves under the Heisenberg Hamiltonian.
In this case, panels (a,b) show that increasing $N$ leads to lower values of $Q$,
in accordance with the fact that $\dim(\rho_1,M=0)$ increases with $N$
(see Table~\ref{tab:groupth_results}).
Hence, the collapse will be investigated more efficiently with larger
systems. Next, we consider the geometry of Fig.~\ref{fig:allgeoms}d
involving $N=12$ spins.
Panel (c) indicates that, for the shorter duration $t_\mathrm{max}=2\, h/J$,
the lowest value of $Q$ is achieved using the Heisenberg Hamiltonian.
However, for the longer duration $t_\mathrm{max}=10\, h/J$, panel (d)
reveals that the Heisenberg Hamiltonian is no longer optimal,
and suggests turning to the more general XXZ Hamiltonian with smaller values of
the ratio $J_z/J$.

\section{ \label{sec:Conclusion}
  Conclusion}

We have theoretically analyzed the time dependence of the measurement
probabilities obtained on an XXZ quantum simulator
comprised of up to $N=2n=12$ interacting particles trapped in a planar geometry
with high spatial symmetry,
namely, point group symmetry $D_{nh}$.
We consider experimentally accessible initial states
which are invariant under all spatial symmetries, i.e.\
which transform under the unit representation $\rho_1$ of
the spatial symmetry group.
Then, the quantum evolution of the component of the $N$--particle
wavefunction with total spin projection $M$ takes place within
the subspace $(\rho_1,M)$ of Sec.~\ref{sec:two_observable_consequences}.
In the case of Sec.~\ref{sec:XXZ_twofoldspinrotation},
where the parity under $Y_\mathrm{spin}$ plays a role,
the relevant subspaces are $(\rho_1,M=0,\mathrm{even}/Y_\mathrm{spin})$
and $(\rho_1,M=0,\mathrm{odd}/Y_\mathrm{spin})$.
The dimensions of these subspaces,
collected in Table~\ref{tab:groupth_results},
determine the qualitative behavior of the
time dependence of the measurement probabilities,
and  are equal to the number of inequivalent measurement results.
These dimensions, calculated using group--theoretical methods,
are characteristic of the spin--point symmetry group of
the Hamiltonian:
our protocol may be understood as a way of determining them experimentally.

The protocol we have put forward
is within experimental reach,
e.g.\ with trapped Rydberg atoms
or polar molecules, owing to recent advances
in trapping techniques \cite{kaufman:NatPhys2021},
in the  quantum simulation of spin Hamiltonians \cite{scholl:PRXQ2022,christakis:Nature2023}
and in the implementation of projective quantum measurements \cite{machu:arXiv2025}.
Our protocol involves initial states that are easy to prepare, and relies on
the measurement scheme whose experimental implementation is the most
straightforward, namely, the simultaneous measurement of the $z$--component
of all $N$ effective spins.

We have highlighted two predictions. Firstly,
the XXZ Hamiltonian is invariant under a twofold
rotation of the $N$ spins about an axis in the horizontal plane.
This yields a conservation law which may be probed efficiently in smaller systems
involving e.g.\ $N=6$ spins (see Sec.~\ref{sec:XXZ_twofoldspinrotation}).
The second highlighted prediction concerns larger systems (comprised of
e.g.\ $N=12$ spins, see Sec.~\ref{sec:Heisenberg_chi_collapse}). 
There, four different qualitative behaviors may be
observed for the time dependence of the measurement probabilities:
these may be constant, or oscillate sinusoidally, or undergo an aperiodic
evolution, or exhibit the collapse of
the component of the initial state with total spin projection $M=0$.
These four behaviors are observed on the same system, prepared in the same
initial state, using the same values
for the parameters $J$ and $J_z$ entering the Hamiltonian of Eq.~\eqref{eq:XXZham}.
Each realization
of the protocol will explore a subspace with given spin projection $M$,
and different subspaces will give rise to different qualitative behaviors
as a function of the protocol duration,
as illustrated in Fig.~\ref{fig:probs_timedep_chit6}.

For larger systems, the number of possible measurement results in the basis
$\mathcal{C}$ grows exponentially with the number of particles $N$.
However, the number of inequivalent measurement
results, whose probabilities are different, is much smaller. For instance,
for $N=12$ particles, there are $\binom{12}{6}=924$ measurement results with
total spin projection $M=0$, among which at most $90$  are inequivalent
(see panels (c,f) of Fig.~\ref{fig:probs_timedep_chit6}).
This is readily exploited by grouping equivalent measurement results
into a single outcome and  plotting their combined probabilities $P_F^M(t)$,
as we have done throughout this paper.

\begin{acknowledgments}
  We acknowledge  stimulating discussions with
  M.~Brune and J.M.~Raimond (LKB, Collège de France, France)
  and R.J.~Papoular (IRAMIS, CEA Saclay, France).
\end{acknowledgments}


\appendix

\bigskip
\centerline{\textbf{APPENDICES}}
\bigskip

The three following Appendices provide additional 
information supporting
our results.  
In App.~\ref{sec:config_numbering}, we summarize the various
orderings used for the $N$--particle configurations
in the basis $\mathcal{C}$.
In App.~\ref{sec:initialstates_parityYspin}, we identify
the $N$--particle states of the form of Eq.~\eqref{eq:def_chi_uv}
whose component with total spin projection $M=0$ is either even or odd
under the operator $Y_\mathrm{spin}$. 
Finally, in App.~\ref{sec:different_probabilities}, we calculate the
number of inequivalent measurement results in the various cases considered
in the main text.

\section{\label{sec:config_numbering}
  Orderings for the $N$--particle states in the basis $\mathcal{C}$}

For a given particle number $N=2n$, the Hilbert space $\mathcal{H}$
has dimension $2^N$.
The basis 
$\mathcal{C}=(\ket{c_f})_{1\leq f\leq 2^{N}}$
of possible measurement results, introduced in Sec.~\ref{sec:protocol},
is comprised of the configurations 
$\ket{\mu_1,\ldots,\mu_{N}}$,
where
the spin at site $A_i$ is in the state
$\ket{\mu_i}=\ket{\uparrow_i^z}$ or
$\ket{\downarrow_i^z}$.
They are labeled by the integer index
$f=1+\sum_{i=1}^{N}(1/2-\mu_i)\, 2^{i-1}$,
where the values $\mu_i=\pm 1/2$
respectively correspond to $\ket{\uparrow_i^z}$
and $\ket{\downarrow_i^z}$.
Hence, $1\leq f\leq 2^N$, with
$\ket{c_1}=\ket{\uparrow^z_1,\ldots,\uparrow^z_{N}}$
and $\ket{c_{2^N}}=\ket{\downarrow^z_1,\ldots,\downarrow^z_{N}}$.

In Sec.~\ref{sec:conservationlaws_measurement}, we sort the states
$\ket{c_f}=\ket{c^M_f}$ in $\mathcal{C}$
in terms of their total spin projection
$M=\sum_{i=1}^{N}\mu_i$.
Hence, $S_z\ket{c_f^M}=\hbar M\ket{c_f^M}$,
with the operator $S_z$ representing the total spin projection along $z$.
For a given  $M$,
there are $f_\mathrm{max}=\binom{N}{n+M}$ states $\ket{c_f^M}$,
labeled with the integer index $f$ such that
$1\leq f\leq f_\mathrm{max}$, ordered by increasing
$\sum_{i=1}^{N}(1/2-\mu_i)\, 2^{i-1}$.
Thus,
$\ket{c_1^{M=0}}=\ket{
  \downarrow^z_1,\ldots,\downarrow^z_{n};
  \uparrow^z_{n+1},\ldots,\uparrow^z_{N}
}$ and
$\ket{c_{f_\mathrm{max}}^{M=0}}=\ket{
  \uparrow^z_1,\ldots,\uparrow^z_{n};
  \downarrow^z_{n+1},\ldots,\downarrow^z_{N}
}$.

Finally, in Sec.~\ref{sec:equiv_results},
among the $\binom{N}{n+M}$
possible measurement results $\ket{c_f^M}$ in the basis $\mathcal{C}$ with total spin projection $M$,
we select a subset of inequivalent states.
We label them with the capital letter `$F$', such that
$\ket{c_F^M}=\ket{c_f^M}$, where $f$ takes the lowest possible value
among the equivalent states $\ket{c_{f'}^M}=U_\phi\ket{c_f^M}$,
all in $\mathcal{C}$.
The number of inequivalent measurement results $\ket{c_F^M}$
depends  on the initial state $\ket{\psi_0}$.
For example, for $N=6$ particles
evolving under $H_\mathrm{XXZ}$,
there are $3$ inequivalent states $\ket{c_F^{M=0}}$
with total spin projection $M=0$ if
$\ket{\psi_0}=\ket{\xi_{\vec{u}}}$ or $\ket{\eta_{\vec{u}}}$,
but there are $6$ of them if
$\ket{\psi_0}=\ket{\chi_{\vec{u},\vec{v}}}$, where
the direction $\vec{v}$ is equal neither to $\vec{u}$
nor to its image under the rotation about $\vec{z}$ through angle $\pi$
(see Fig.~\ref{fig:M0Yspin}).

\section{\label{sec:initialstates_parityYspin}
  Initial states with a well--defined parity
  under $Y^\mathrm{spin}$
}

In this section, we identify all states $\ket{\chi}=\ket{\chi_{\vec{u},\vec{v}}}$,
of the form of Eq.~\eqref{eq:def_chi_uv}, whose component $\ket{\chi^{M=0}}$
with total spin projection $M=0$ is an eigenstate of the operator
$Y_\mathrm{spin}$ of Eq.~\eqref{eq:Yspin}.
Thus, we seek states $\ket{\chi}$ such that
$Y_\mathrm{spin}\ket{\chi^{M=0}}=\epsilon\ket{\chi^{M=0}}$,
where the eigenvalue $\epsilon=\pm 1$ determines
the even or odd
parity of $\ket{\chi^{M=0}}$ with respect to the operator
$Y_\mathrm{spin}$.

We consider the geometry involving $N=2n$ spins.
We write  the single--particle state $\ket{\uparrow^{\vec{u}}}$ used for all $n$ sites
on the outer ring as
$\ket{\uparrow^{\vec{u}}}=a_{\vec{u}}\ket{\uparrow^z}+b_{\vec{u}}\ket{\downarrow^z}$,
where $a_{\vec{u}}=\cos(\theta_{\vec{u}}/2)\, e^{-i\phi_{\vec{u}}/2}$,
$b_{\vec{u}}=\sin(\theta_{\vec{u}}/2)\, e^{+i\phi_{\vec{u}}/2}$,
and the angles $(\theta_{\vec{u}},\phi_{\vec{u}})$
are the spherical coordinates of the unit vector $\vec{u}$
on the Bloch sphere.
Similarly, we
write  the single--particle state $\ket{\uparrow^{\vec{v}}}$ used for all $n$ sites
on the inner ring as
$\ket{\uparrow^{\vec{v}}}=a_{\vec{v}}\ket{\uparrow^z}+b_{\vec{v}}\ket{\downarrow^z}$,
where
$a_{\vec{v}}=\cos(\theta_{\vec{v}}/2)\, e^{-i\phi_{\vec{v}}/2}$,
$b_{\vec{v}}=\sin(\theta_{\vec{v}}/2)\,  e^{+i\phi_{\vec{v}}/2}$,
and the angles $(\theta_{\vec{v}},\phi_{\vec{v}})$
are the spherical coordinates of the unit vector $\vec{v}$
on the Bloch sphere. No solution is found if one or more of the four complex numbers
$a_{\vec{u}}$, $b_{\vec{u}}$, $a_{\vec{v}}$, $b_{\vec{v}}$
is zero, hence, we assume that they are all non--zero.
The $M=0$ component of the $N$--particle state $\ket{\chi}$ reads:
\begin{equation}
  \label{eq:chiM0_gamma}
  \ket{\chi^{M=0}}=\sum_{n_{O\uparrow}=0}^n
  \left( a_{\vec{u}} b_{\vec{v}}\right)^{n_{O\uparrow}}
  \left( b_{\vec{u}} a_{\vec{v}}\right)^{n-n_{O\uparrow}}
  \ket{\gamma_{n_{O\uparrow}}}
  \ ,
\end{equation}
where the (non--normalized) $N$--particle state $ \ket{\gamma_{n_{O\uparrow}}}$
is the sum of all states $\ket{c_f^{M=0}}$ in the basis $\mathcal{C}$ whose
total spin projection is $M=0$, and which have exactly $n_{O\uparrow}$ spins
on the outer (`O') ring in the state $\ket{\uparrow^z}$. The operator $Y_\mathrm{spin}$
maps $\ket{\chi^{M=0}}$ onto:
\begin{equation}
  \label{eq:YspinchiM0_gamma}
  Y_\mathrm{spin}\ket{\chi^{M=0}}=\sum_{n_{O\uparrow}=0}^n
  \left( -a_{\vec{u}} b_{\vec{v}}\right)^{n-n_{O\uparrow}}
  \left( -b_{\vec{u}} a_{\vec{v}}\right)^{n_{O\uparrow}}
  \ket{\gamma_{n_{O\uparrow}}}
  \ .
\end{equation}
We introduce the complex number
$z=(a_{\vec{u}}b_{\vec{v}})/(b_{\vec{u}} a_{\vec{v}})$.
Owing to Eqs.~\eqref{eq:chiM0_gamma} and \eqref{eq:YspinchiM0_gamma}, 
the relation $Y_\mathrm{spin}\ket{\chi^{M=0}}=\epsilon\ket{\chi^{M=0}}$
requires $(-z)^{n-2n_{O\uparrow}}=\epsilon$ for any integer $n_{O\uparrow}$
such that $0\leq n_{O\uparrow}\leq n$. Hence, $z=\pm 1$.
The case $z=+1$ yields the state
$\ket{\chi}=\ket{\xi_{\vec{u}}}=\ket{\chi_{\vec{u},\vec{u}}}$ introduced
in Sec.~\ref{sec:initstates},
and the corresponding eigenvalue
$\epsilon=(-1)^n$ depends on the parity of $n$.
The case $z=-1$ yields
the state $\ket{\chi}=\ket{\eta_{\vec{u}}}=\ket{\chi_{\vec{u},\vec{u}'}}$,
where the unit vector $\vec{u}'$ on the Bloch sphere
is the image of $\vec{u}$ under the rotation through
angle $\pi$ about the axis $\vec{z}$. The corresponding eigenvalue
is $\epsilon=+1$
for all values of $n$.

To sum up, if $n$ is odd, the
$M=0$ components $\ket{\xi^{M=0}_{\vec{u}}}$ and $\ket{\eta^{M=0}_{\vec{u}}}$
of the states $\ket{\xi_{\vec{u}}}$
and $\ket{\eta_{\vec{u}}}$ are respectively odd and even under
the operator $Y_\mathrm{spin}$, 
as illustrated in
Fig.~\ref{fig:M0Yspin} for $n=3$.
By contrast, if $n$ is even, both $\ket{\xi^{M=0}_{\vec{u}}}$ and
$\ket{\eta^{M=0}_{\vec{u}}}$ are even with respect to $Y_\mathrm{spin}$,
and there is no state $\ket{\chi_{\vec{u},\vec{v}}}$ of the form of Eq.~\eqref{eq:def_chi_uv}
whose $M=0$ component is odd under $Y_\mathrm{spin}$.

\section{\label{sec:different_probabilities}
  Numbers of inequivalent measurement results
}

\subsection{ \label{sec:permrep:totalspinprojection}
  Role of the spatial symmetries}

In this section, we derive the number of inequivalent measurement results
$\ket{c_F^M}$
for choices of the Hamiltonian ($H_\mathrm{H}$ or $H_\mathrm{XXZ}$)
and the initial state $\ket{\psi_0}$ such that
the only relevant symmetries are  \textit{(i)}
the spatial symmetries in the group $D_{nh}$
and \textit{(ii)} the conservation of the total spin projection $S_z$.

We consider the geometry involving $N=2n$ particles.
For a given total spin projection $M$,
we call $\mathcal{H}^M$ the subspace of the Hilbert space $\mathcal{H}$
comprised of all $N$--particle states with total spin projection $M$.
It is spanned by the $\binom{N}{n+M}$ states $\ket{c_f^M}$
in the basis $\mathcal{C}$ with total spin projection $M$.

The unitary operators $U_\phi$
acting on the Hilbert space $\mathcal{H}$,
introduced in Sec.~\ref{sec:sym_space_spin}, all commute with
the total spin projection operator $S_z$.
Hence, they leave the subspace $\mathcal{H}^M$ invariant.
Thus, they make up a (reducible) representation
$\mathcal{R}^M$, acting on $\mathcal{H}^M$,
of the spatial symmetry group $G^\mathrm{spatial}=D_{nh}$.
All operators $U_\phi$ map each state $\ket{c^M_f}$
in the basis $\mathcal{C}$ onto a state $\ket{c^M_{f'}}$, also in the
basis $\mathcal{C}$ (rather than onto a linear combination of basis states).
Linear representations satisfying this property are known
as permutation representations \cite[Sec.~1.2]{serre:Springer1977}.

Two possible measurement results $\ket{c^M_f}$ and $\ket{c^M_{f'}}$
are `equivalent' if one is mapped onto the other by some 
symmetry operator $U_\phi$, namely, $\ket{c^M_{f'}}=U_\phi\ket{c^M_f}$
(see Sec.~\ref{sec:equiv_results}).
Hence, the number 
of inequivalent measurement results is the number
of different sets $\{U_\phi\ket{c^M_f}\}$,
called `orbits', 
obtained by applying all 
operators $U_\phi$ to each state $\ket{c^M_f}$ in $\mathcal{C}$.
Owing to a known property of permutation
representations \citep[Sec.~2.3]{serre:Springer1977}, this number
is equal to the dimension
of the subspace of $\mathcal{H}^M$ transforming under $\mathcal{R}^M$
according to the unit representation of $D_{nh}$, i.e.\ to the dimension
$\dim(\rho_1,M)$,
as stated in Sec.~\ref{sec:equiv_results}.

\subsection{ \label{sec:permrep:twofoldspinrotation}
  Case of $M=0$: role of the two--fold rotations of the $N$ spins}
We now  derive the number of inequivalent measurement results
$\ket{c_F^{M=0}}$
with total spin projection $M=0$
for choices of the Hamiltonian
and initial state such that,
in addition to the symmetries accounted for in Sec.~\ref{sec:permrep:totalspinprojection},
the spin rotations through angle $\pi$ about any horizontal axis also play a role. 

The group of spin symmetries $G^\mathrm{spin}=D_{\infty h}$
is comprised of the products of all rotations about the
axis $\vec{z}$ through any angle, all rotations
through angle $\pi$
about any axis in the horizontal plane $(Oxy)$, and inversion.
The unitary operators $U_g$ of Sec.~\ref{sec:sym_space_spin} make up
a representation of $G^\mathrm{spin}$ acting on the Hilbert space $\mathcal{H}$.
This representation is single--valued, because
the considered system is comprised of an even number $N=2n$ of 
spins--$1/2$ \cite[\S 99]{landau3:BH1977}.
Inversion acts as the identity
because spins are pseudovectors \cite[Sec.~15.10]{messiah2:NorthHolland1962}.

The subspace
$\mathcal{H}^{M=0}$
is invariant under all operators $U_g$.
Within it,
all spin rotations $C^{\vec{z}}_{\phi}$
about the axis $\vec{z}$ through  angle $\phi$ act as the
identity \cite[Sec.~XIII.20]{messiah2:NorthHolland1962}.
Moreover, the spin rotation  $C^{\vec{e}}_{\pi}$
through angle $\pi$ about the horizontal axis with polar angle $\phi$,
namely, $\vec{e}=(\cos\phi,\sin\phi,0)$,
satisfies the geometric relation
$C_\pi^{\vec{e}}=C_\phi^{\vec{z}} C_\pi^{\vec{x}} C_{-\phi}^{\vec{z}}$.
Therefore, all spin rotations $C^{\vec{e}}_{\pi}$ act on $\mathcal{H}^{M=0}$
as the same operator. Hence, it is sufficient to account for a single such rotation,
say $C_\pi^{\vec{y}}$. 
Thus, the behavior of the
states
in $\mathcal{H}^{M=0}$ under all spatial and spin symmetries is fully determined
by the  group
$G_0=D_{nh}^\mathrm{spatial}\times
\{ 1,C_\pi^{\vec{y}}\}^\mathrm{spin}$, which is
the direct product of the spatial symmetry group $D_{nh}$ with
a group 
comprised of two spin symmetries.
The group $G_0$ is a finite subgroup of the full spin--point group
$G$ of Sec.~\ref{sec:sym_space_spin}.

The operators $U_\phi$ and $U_g$ of Sec.~\ref{sec:sym_space_spin}
yield a reducible representation $\mathcal{R}_0$
of the group $G_0$ acting on the subspace $\mathcal{H}^{M=0}$.
In particular, the spin rotation $C_\pi^{\vec{y}}$ acts as
$Y_\mathrm{spin}^{M=0}$,
where:
\begin{equation}
  \label{eq:Yspin_M=0}
  Y_\mathrm{spin}^{M=0} = (-1)^n F_\mathrm{spin}^{M=0}
  \ .
\end{equation}
In Eq.~\eqref{eq:Yspin_M=0}, $Y_\mathrm{spin}^{M=0}$
and $F_\mathrm{spin}^{M=0}$
are the 
restrictions to the
subspace $\mathcal{H}^{M=0}$ of the operator
$Y_{\mathrm{spin}}$ 
representing the spin rotation $C_\pi^{\vec{y}}$,
and of the operator $F_\mathrm{spin}=\sigma_1^x\ldots\sigma_N^x$
flipping the projection along $z$ of each individual spin
($\ket{\uparrow_i^z}$ and $\ket{\downarrow_i^z}$
are respectively mapped onto $\ket{\downarrow_i^z}$ and $\ket{\uparrow_i^z}$).

We introduce 
the permutation representation $\mathcal{R}_0^+$
of $G_0$ acting on $\mathcal{H}^{M=0}$ defined as follows: all spatial
symmetries act as in the representation $\mathcal{R}_0$,
but the spin rotation $C_\pi^{y,\mathrm{spin}}$ acts
as $+F_\mathrm{spin}^{M=0}$. 
The property of permutation representations already used in
Appendix~\ref{sec:permrep:totalspinprojection} above now yields the following result.
The number of
inequivalent measurement results $\ket{c_F^{M=0}}$ with total spin projection $M=0$ 
is equal to the dimension of the subspace of $\mathcal{H}^{M=0}$
comprised of the states transforming under $\mathcal{R}_0^+$
according to the unit representation of $G_0$,
namely, the states invariant under all spatial symmetries and under $F_\mathrm{spin}$.

If $n$ is even, Eq.~\eqref{eq:Yspin_M=0} shows that
the representations $\mathcal{R}_0$ and $\mathcal{R}_0^+$ coincide.
Then, the number of inequivalent measurement results $\ket{c_F^{M=0}}$ is
$\dim(\rho_1,M=0,\text{even/$Y_\mathrm{spin}$})$. We have confirmed
this prediction numerically
for the initial states $\ket{\xi_{\vec{x}}}$ and $\ket{\eta_{\vec{x}}}$,
in the cases of the geometries of Figs.~\ref{fig:allgeoms}b and
\ref{fig:allgeoms}d,
which respectively
involve $N=2n=8$ and $12$ atoms
(for these geometries,
the components $\ket{\xi_{\vec{x}}^{M=0}}$ and $\ket{\eta_{\vec{x}}^{M=0}}$
are both even under $Y_\mathrm{spin}$:
see Appendix \ref{sec:initialstates_parityYspin} above).

If $n$ is odd, $F_\mathrm{spin}^{M=0}=-Y_\mathrm{spin}^{M=0}$, so that
the number of inequivalent measurement
results $\ket{c_F^{M=0}}$ is equal to
$\dim(\rho_1,M=0,\text{odd/$Y_\mathrm{spin}$})$,
which is also equal to $\dim(\rho_1,M=0,\text{even/$Y_\mathrm{spin}$})$
(see Appendix~\ref{sec:compare_dim_M0even_M0odd} below).
This prediction is in full agreement with our numerical
results, illustrated in Fig.~\ref{fig:M0Yspin}e
for the initial states $\ket{\xi_{\vec{x}}}$
and $\ket{\eta_{\vec{x}}}$ involving $N=2n=6$ atoms
(geometry of Fig.~\ref{fig:allgeoms}a), which are
respectively eigenstates of 
$Y_\mathrm{spin}$ with eigenvalues $-1$ and $+1$.

\subsection{ \label{sec:compare_dim_M0even_M0odd}
  Comparing the
  dimensions of the subspaces $(\rho_1,M=0,\mathrm{even}/Y^\mathrm{spin})$
  and $(\rho_1,M=0,\mathrm{odd}/Y^\mathrm{spin})$
}

The dimensions
$d_\mathrm{even}= \dim(\rho_1,M=0,\mathrm{even}/F_\mathrm{spin})$,
$d_\mathrm{odd}=\dim(\rho_1,M=0,\mathrm{odd}/F_\mathrm{spin})$,
and $d=\dim(\rho_1,M=0)$,
satisfy $d_\mathrm{even}+d_\mathrm{odd}=d$.
We further relate the dimensions 
$d$ and $d_\mathrm{even}$ by interpreting them as the numbers
of orbits for two different permutation representations,
both acting on the subspace
$\mathcal{H}^{M=0}$ of $N$--particle states with total spin projection $M=0$.
The first one, $\mathcal{R}^{M=0}$, introduced in
Appendix \ref{sec:permrep:totalspinprojection} above, is a representation of
the spatial symmetry group $D_{nh}$.
Its number of orbits is $d=\dim(\rho_1,M=0)$.
The second one, $\mathcal{R}_0^+$,
introduced in Appendix \ref{sec:permrep:twofoldspinrotation},
is a representation of the subgroup $G_0$ of the spin--point group $G$.
Its number of orbits is
$d_\mathrm{even}=\dim(\rho_1,M=0,\mathrm{even}/F_\mathrm{spin})$,
the parity under $F_\mathrm{spin}$ being determined
by the parity under $Y_\mathrm{spin}$
through Eq.~\eqref{eq:Yspin_M=0}.

We consider the orbit $\Omega$, under the representation $\mathcal{R}_0^+$,
of the configuration $\ket{c_{f_0}}=\ket{c_{f_0}^{M=0}}$
with total spin projection $M=0$.
It is comprised of the distinct elements among
$\{U_\phi\ket{c_{f_0}}\}$ and $\{F_\mathrm{spin}U_\phi\ket{c_{f_0}}\}$,
for all spatial symmetries $\phi$ in $D_{nh}$, the operators $U_\phi$
being defined in Sec.~\ref{sec:sym_space_spin}.
There are two cases:
\begin{enumerate}[label=(\alph*)]
\item If $F_\mathrm{spin}\ket{c_{f_0}}=U_{\phi_0}\ket{c_{f_0}}$
  for some spatial symmetry $\phi_0$, then all elements in $\Omega$
  may be written as $U_\phi\ket{c_{f_0}}$. Thus, $\Omega$ is also
  an orbit 
  under the representation $\mathcal{R}^{M=0}$.
\item If $F_\mathrm{spin}\ket{c_{f_0}}\neq U_\phi \ket{c_{f_0}}$ 
  for all spatial symmetries $\phi$, $\Omega$
  yields two different orbits under the representation $\mathcal{R}^{M=0}$,
  namely, the sets $\{U_\phi\ket{c_{f_0}}\}$ and
  $\{F_\mathrm{spin}U_\phi\ket{c_{f_0}}\}$.
\end{enumerate}
We call $\lambda_a$ and $\lambda_b$ the numbers of orbits of $\mathcal{R}_0^+$
satisfying cases (a) and (b), respectively.
Thus, $d_\mathrm{even}$ and $d$ satisfy:
\begin{equation}
  d_\mathrm{even}=\lambda_a+\lambda_b
  \text{\quad and \quad}
  d=\lambda_a+2\lambda_b
  \ .
\end{equation}
Hence, the difference $d_\mathrm{even}-d_\mathrm{odd}=2d_\mathrm{even}-d=\lambda_a$.

\emph{Distinction between even and odd values of $n$ ---}
We write $\ket{c_{f_0}}=\ket{\mu_1,\ldots,\mu_N}$ with $N=2n$.
We introduce the spin projections
$M_O=\mu_1+\ldots+\mu_n$ and $M_I=\mu_{n+1}+\ldots+\mu_{N}$ on the
outer (`$O$') and inner (`$I$') rings, with $\mu_i=\pm 1/2$
according to whether $\ket{\mu_i}=\ket{\uparrow_i^z}$ or $\ket{\downarrow_i^z}$.
For any spatial symmetry $\phi$, the state $U_\phi\ket{c_{f_0}}$
has the same spin projections $M_O$ and $M_I$, but the state
$F_\mathrm{spin}\ket{c_{f_0}}$ has the spin projections $-M_O$ and $-M_I$.
Therefore, case (a) requires 
$M_O=M_I=0$, which is only possible if $n$ is even.
Thus, $\lambda_a=0$ for odd values of $n$.
To conclude,
$d_\mathrm{even}=d_\mathrm{odd}$ if $n$ is odd,
and $d_\mathrm{even} > d_\mathrm{odd}$ if $n$ is even.
These results are confirmed by our explicit calculations
for $n=3$, $4$, $5$, and $6$, summarized in Table \ref{tab:groupth_results}
of the main text.


%

\end{document}